%% file: main.tex
\documentclass[sigconf, nonacm, table]{acmart}
\usepackage{xcolor}
\usepackage{popets}

\setcopyright{popets}
\copyrightyear{YYYY}
\acmYear{YYYY}
\acmVolume{YYYY}
\acmNumber{X}
\acmDOI{XXXXXXX.XXXXXXX}
\acmISBN{}
\acmConference{Proceedings on Privacy Enhancing Technologies}

\usepackage{listings}
\usepackage{tikz}
\usepackage{soul}
\usepackage{nicematrix}
\usepackage[normalem]{ulem}
\usepackage{diagbox}
\usepackage{multirow}
\usepackage{graphicx}
\usepackage{array}
\usepackage{paralist}
\usepackage{enumitem}
\usepackage{wrapfig}
\usepackage{cleveref}

\newif\ifdraft % set to false to remove comments and todos
\draftfalse

\definecolor{LightRed}{rgb}{1,0.6,0.6}
\definecolor{LightRedbis}{rgb}{0.9,0.9,0.9}
\definecolor{mygreen}{rgb}{0.1, 0.5, 0.1}
\definecolor{myblue}{RGB}{220, 231, 250}

\usepackage{array} % Add this to your preamble if not already included

\definecolor{lightred}{RGB}{255,200,200}
\definecolor{lightorange}{RGB}{255,255,200}
\definecolor{lightgreen}{RGB}{200,255,200}

\definecolor{textred}{RGB}{255,0,0}
\definecolor{textorange}{RGB}{255,140,0}
\definecolor{textgreen}{RGB}{0,128,0}
\newcommand{\red}[1]{\textcolor{textred}{#1}}
\newcommand{\orange}[1]{\textcolor{textorange}{#1}}
\newcommand{\green}[1]{\textcolor{blue}{#1}}

\newcommand{\viando}[1]{\textcolor{red}{#1}}
\newcommand{\azure}[1]{\textcolor{blue}{#1}}
\newcommand{\aws}[1]{\textcolor{orange}{#1}}
\newcommand{\local}[1]{\textcolor{olive}{#1}}
\newcommand{\old}[1]{\textcolor{gray}{#1}}

\newcommand{\Epsi}{\mathcal{E}}

\newcommand{\Att}{$\mathcal{A}$}

\definecolor{customblue}{RGB}{0, 0, 255}
\definecolor{BLUE}{RGB}{0, 0, 255}
\newcommand{\change}[1]{#1}

\usepackage{enumitem}
\setlist[itemize]{leftmargin=*}

\PassOptionsToPackage{noabbrev,capitalize}{cleverref}

\lstdefinelanguage[RISC-V]{Assembler}
{
  alsoletter={.}, % allow dots in keywords
  alsodigit={0x}, % hex numbers are numbers too!
  morekeywords=[1]{ % instructions
    lb, lh, lw, lbu, lhu,
    sb, sh, sw,
    sll, slli, srl, srli, sra, srai,
    add, addi, sub, lui, auipc,
    xor, xori, or, ori, and, andi,
    slt, slti, sltu, sltiu,
    beq, bne, blt, bge, bltu, bgeu,
    j, jr, jal, jalr, ret,
    scall, break, nop, csrwi, mv, ld, sd
  },
  morekeywords=[2]{ % sections of our code and other directives
    .align, .ascii, .asciiz, .byte, .data, .double, .extern,
    .float, .globl, .half, .kdata, .ktext, .set, .space, .text, .word
  },
  morekeywords=[3]{ % registers
    zero, ra, sp, gp, tp, s0, fp,
    t0, t1, t2, t3, t4, t5, t6,
    s1, s2, s3, s4, s5, s6, s7, s8, s9, s10, s11,
    a0, a1, a2, a3, a4, a5, a6, a7,
    ft0, ft1, ft2, ft3, ft4, ft5, ft6, ft7,
    fs0, fs1, fs2, fs3, fs4, fs5, fs6, fs7, fs8, fs9, fs10, fs11,
    fa0, fa1, fa2, fa3, fa4, fa5, fa6, fa7
  },
  morecomment=[l]{;},   % mark ; as line comment start
  morecomment=[l]{//},   % mark ; as line comment start
  morecomment=[l]{\#},  % as well as # (even though it is unconventional)
  morestring=[b]",      % mark " as string start/end
  morestring=[b]'       % also mark ' as string start/end
}

\makeatletter
\newcommand{\code}[1]{\protect\lstinline[language={[RISC-V]Assembler},basicstyle=\ttfamily]|#1|}
\makeatother

\definecolor{mauve}{rgb}{0.58,0,0.82}
\lstdefinestyle{myCstyle}{
  language=C,
  frame=none,
  basicstyle=\tiny\ttfamily,
  keywordstyle=\color{blue},
  keywordstyle=[2]\color{red},
  stringstyle=\color{red},
  commentstyle=\color{green},
  morekeywords=[2]{BURST_ON, BURST_OFF},
  showstringspaces=false
}

\begin{document}
\title{Teaching an Old Dog New Tricks: Verifiable FHE Using Commodity Hardware}

%% \citi{} and \state{} are fully optional.
%% You do not need to provide values for the \city{}, \state{}, and \country{} commands 
%% within the \author{} commands, but YOU MUST leave the blank commands in the file.

\author{Jules Drean}
\affiliation{\institution{MIT CSAIL}
\city{}
\state{}
\country{}}

\author{Fisher Jepsen}
\affiliation{\institution{MIT CSAIL}
\city{}
\state{}
\country{}}

\author{G. Edward Suh}
\affiliation{\institution{NVIDIA / Cornell University}
\city{}
\state{}
\country{}}

\author{Srinivas Devadas}
\affiliation{\institution{MIT CSAIL}
\city{}
\state{}
\country{}}

\author{Aamer Jaleel}
\affiliation{\institution{NVIDIA}
\city{}
\state{}
\country{}}

\author{Gururaj Saileshwar}
\affiliation{\institution{University of Toronto}
\city{}
\state{}
\country{}}

\renewcommand{\shortauthors}{Drean et al.}

\input{sections/abstract}


\maketitle

\input{sections/introduction}
\input{sections/threat_model}
\input{sections/background}
\input{sections/overview}
\input{sections/remote_attesation}
\input{sections/security_annalysis}
\input{sections/applications}
\input{sections/implementation}
\input{sections/evaluation}
\input{sections/relatedwork}
\input{sections/conclusion}

\begin{acks}
The authors thank Charly Castes, Adrien Ghosn, Neelu S. Kalani, and the authors of Tyche \cite{castes2023creating} for providing early access to their code and for their invaluable assistance in setting up Tyche on our hardware and adapting the platform to our needs.
We thank Sacha Servan-Schreiber for helpful discussions on fully homomorphic encryption and useful feedback.
The authors used Claude to revise the text in most of this paper to correct for typos, grammatical errors, and awkward phrasing.
This research was supported in part by NSF contracts 2330065, 2115587 and 1955270, and Natural Sciences and Engineering Research Council of Canada (NSERC) under funding reference number RGPIN-2023-04796, and an NSERC-CSE Research Communities Grant under funding reference number ALLRP-588144-23.
Any research, opinions, or positions expressed in this work are solely those of the authors and do not represent the official views of the Communications Security Establishment Canada or the Government of Canada. 
\end{acks}

\bibliographystyle{ACM-Reference-Format}
\bibliography{sample-base}

\input{sections/appendix}

\end{document}
\endinput
%%
%% End of file `sample-sigconf.tex'.

%% file: sections/abstract.tex
\begin{abstract}
We present Argos, a simple approach for adding verifiability to fully homomorphic encryption (FHE) schemes using trusted hardware. 
Traditional approaches to verifiable FHE require expensive cryptographic proofs, which incur an overhead of up to seven orders of magnitude \emph{on top of FHE}, making them impractical.

With Argos, we show that trusted hardware can be securely used to provide verifiability for FHE computations, with minimal overhead relative to the baseline FHE computation.
An important contribution of Argos is showing that the major security pitfall associated with trusted hardware, \emph{microarchitectural} side channels, can be completely mitigated by excluding any secrets from the CPU and the memory hierarchy.
This is made possible by focusing on building a platform that only enforces program and data \emph{integrity} and \emph{not} confidentiality (which is sufficient for verifiable FHE, since all data remain encrypted at all times).
All secrets related to the attestation mechanism are kept in a separate coprocessor (e.g., a TPM)---inaccessible to any software-based attacker.
Relying on a discrete TPM typically incurs significant performance overhead, which is why (insecure) software-based TPMs are used in practice. 
As a second contribution, we show that for FHE applications, the attestation protocol can be adapted to only incur a fixed cost.

Argos requires no dedicated hardware extensions and is supported on commodity processors from 2008 onward.
Our prototype implementation introduces 3\% overhead for FHE evaluation, and 8\% for more complex protocols.
In particular, we show that Argos can be used for real-world applications of FHE, such as private information retrieval (PIR) and private set intersection (PSI), where providing verifiability is imperative. 
By demonstrating how to combine cryptography with trusted hardware, Argos paves the way for widespread deployment of FHE-based protocols beyond the semi-honest setting, without the overhead of cryptographic proofs. 
\end{abstract}

%% file: sections/introduction.tex
\section{Introduction}

Fully homomorphic encryption (FHE)~\cite{gentry2009fully, chillotti2020tfhe, ducas2015fhew, chen2017simple} makes it possible to evaluate any logical circuit directly on encrypted data.
In the client-server setting, it can be used to build other powerful cryptographic primitives such as private set intersection (PSI)~\cite{cong2021labeled, angel2018pir, menon2022spiral, henzinger2023one, menon2024ypir}, private information retrieval (PIR)~\cite{cong2021labeled, chen2017fast, chen2018labeled, kiss2017private} or multi-party computation (MPC)~\cite{asharov2012multiparty, dahl2023noah}.
FHE also has many practical use cases including private contact discovery~\cite{signalcontact}, private smart contracts~\cite{smart2023practical}, or private inference~\cite{natarajan2021chex}.
Reducing performance overhead (currently 3 to 7 orders of magnitudes over non-private computation) has been the main focus of the last decade of research, but other limitations of FHE are now becoming more relevant.

At the forefront of these issues, the fact that FHE-schemes are not secure when considering a malicious evaluator, or that existing FHE schemes lack notions of integrity, are significant limitations for real-world deployment~\cite{atapoor2024verifiable, viand2023verifiable, chase2017security}.
First, if a malicious server is able to supply malformed ciphertext for a client to decrypt, it can mount key recovery attacks~\cite{guo2024key, chenal2015key} and break all security (if the attacker recovers the private key, it can now decrypt the client's private requests).
Second, if FHE enables delegation of computation on confidential data to an untrusted party, existing constructions cannot help the user \emph{verify} that the correct function was evaluated.
In other words, FHE schemes always assume honest-but-curious attackers.
In the case of smart contracts, for instance, a malicious evaluator can completely change the functionality of the contract, with the option to even reveal confidential inputs and violate privacy.
A straightforward solution is to add a layer of verifiable computation on top of the FHE scheme.
Unfortunately, existing solutions suffer from major limitations.
Cryptographic proofs incur an overhead of 4 to 6 orders of magnitude on top of FHE~\cite{viand2023verifiable}.
Another solution is to replicate the evaluation across several non-colluding parties, but distributed-trust setups are often impractical in the real world~\cite{lindell2023deployment}.

Trusted execution environments (TEEs) or secure enclaves~\cite{costan2016intel, cheng2024intel, sev2020strengthening,pinto2019demystifying} have also been considered as a potential solution \cite{viand2023verifiable, natarajan2021chex, kumar2020cryptflow}.
These technologies implement remote attestation, providing processors the capability to attest the code they are running, even in the presence of a privileged software attacker.
TEEs offer better performance than cryptographic proofs, but come with weaker security guarantees: an attacker does not need to break a cryptographic assumption to forge a proof. Instead, the attacker needs to hack the platform or steal the attestation key.
Still, these platforms initially seemed to offer a reasonable compromise.
Unfortunately, a long series of published attacks (the SoK of Li et al.~\cite{li2024sok} covers 43 of them) quickly revealed that they were extensively vulnerable to microarchitectural side channels and transient execution attacks \cite{gotzfried2017cache,van2020sgaxe,van2018foreshadow, chen2019sgxpectre, xu2024trustzonetunnel, ryan2019hardware}.
These powerful attacks can be mounted in software and exploit microarchitectural structures and speculative mechanisms present in modern processors to extract secrets from trusted environments, not only violating confidentiality of the protected programs but also extracting keys used for attestation.
New microarchitectural side channels are discovered every year~\cite{moghimi2023downfall, wang2022hertzbleed} and the current consensus is that it will take many years of research to build software and hardware defense mechanisms that can efficiently eliminate these threats from modern processors.
On top of this existential threat, the availability of TEEs is also limited.
Modern platforms require dedicated hardware and are only available on server-grade processors, limiting adoption.
\\

\textbf{This Work.} We present Argos, the first enclave platform specifically designed to build maliciously-secure and verifiable FHE.
As observed in previous work~\cite{kumar2020cryptflow, natarajan2021chex, viand2023verifiable, tramer2017sealed}, for use cases where all application data is encrypted, such as in FHE, the only secrets exposed on the server are those used for attestation.
We go further and show how this is a unique opportunity to rethink TEE architecture and design \emph{integrity-only} enclaves that focus on data and code integrity \emph{without} offering any confidentiality guarantee. 
Lessening the requirements for enclaves gives us the opportunity to
1) completely eliminate microarchitectural side channels by secluding all secret key material in a physically-separated coprocessor (such as a trusted platform module or TPM),
2) use a hypervisor-based TEE architecture that offers better performance and hardware compatibility, and
3) build a simplified attestation scheme that makes it possible to easily prove the security of our construction and amortize the cost of relying on a slow coprocessor for key operations.

\textbf{Eliminating Microarchitectural Side Channels.}
An important observation we make is that side-channel attacks are ``read-only'' gadgets and can be eliminated simply by excluding any secrets from the CPU and the memory hierarchy (see \Cref{fig:computing_stack}).
Software-mounted side channels require an attacker program to share hardware resources with the victim programs in order to extract secrets.
If all secrets are secluded on a separate chip, such as a physical trusted platform module (TPM) or a secure coprocessor~\cite{IntelCSME, appleplatformsecurity, johnson2018titan}, no attacker program can ever share microarchitectural resources with the cryptographic algorithm and extract secrets.
Leveraging this insight, we design Argos to store all attestation secrets inside a discrete TPM and delegate all sensitive operations to outside of the CPU.
This simple principle makes Argos secure \emph{by construction} against side-channels.
The only (side) channel left between the secluded cryptographic keys and a potential attacker is completion time~\cite{moghimi2020tpm} which is addressed through using constant-time cryptography in the TPM (see our threat model in \Cref{sec:threat_model}).

\textbf{Rehabilitating Hypervisor-based TEEs.}
Argos relies on a hypervisor-based TEE design.
A security monitor runs with hypervisor privilege and provides code integrity and isolation to enclave programs.
Such TEEs were introduced by TrustVisor~\cite{mccune2010trustvisor} and are still widely used by cloud vendors to implement confidential computing~\cite{ApplePrivateCloud}.
These architectures are usually considered less secure than dedicated hardware platforms such as Intel SGX or AMD-SEV as they do not encrypt main memory and offer no security against Coldboot~\cite{halderman2009lest}, one of the most practical physical attacks (see \Cref{tab:tee_platforms}).
In our context, this is not an issue as no secrets are ever exposed to main memory.
As a result, despite our strong threat model, Argos can adapt this architecture to bring compatibility with most commodity hardware.
In terms of performance, existing platforms typically rely on virtual TPMs~\cite{raj2016ftpm}---one per execution environment---that are endorsed by the root TPM but executed on CPU~\cite{AzureVirtualTPM, AWSNitroTPM}.
While this approach avoids the performance bottleneck of a single physical TPM~\cite{mccune2010trustvisor}, it makes them vulnerable to microarchitectural side-channels.
In contrast, Argos prefers a secure-by-design approach, centered on the discrete TPM. 

\textbf{Performance Optimization With a Discrete Chip.}
Relying on a single discrete TPM can have a significant impact on performance~\cite{mccune2008flicker}.
Nevertheless, we can still virtualize most TPM resources in the security monitor, trusted code that runs at hypervisor privilege level, as measurement (e.g., hashing) does not require any secret manipulation.
Our use case also excludes the use of the TPM ``sealed'' secret storage.
That means we only use the TPM as a ``signing oracle''.
This can still be expensive, especially when considering interactive cryptographic protocols that require repeated and attested message exchanges.

\textbf{Simple and Efficient Attestation Scheme.}
We show that for our FHE applications, we can build a simple attestation scheme that only ever requires one TPM signature.
Unlike in other enclave platforms, the remote user does not need to see any attestation proof before sending its sensitive data, as it will always stay encrypted.
This also applies to intermediate protocol messages that do not need to be attested but simply added to a transcript whose hash is extended in the security monitor.
At the end of the FHE computation (or more precisely, before any FHE ciphertext needs to be decrypted), the transcript's hash will be signed using the TPM.
We show how our simplified transcript-based attestation can be model as a proof system, and show that it is sufficient to achieve malicious security needed for verifiable FHE.
We also show how Argos can be extended to support batched verifiable FHE.
As a result, the overhead of using a physical TPM becomes a fixed cost, and performance becomes similar to that of a virtual TPM, while enforcing security-with-abort in the presence of a malicious attacker~\cite{goldreich2019play}.

\textbf{Extending Argos to FHE-Based Applications.}
Circuit-level security sometimes differs from application-level security.
In many FHE-based applications, a malicious server can provide corrupted inputs and gain some information from how the client behaves following decryption.
This can have devastating downstream security implications. 
We show how Argos can easily be extended to support complex protocols in the malicious setting, such as authenticated PIR or authenticated PSI with almost no overhead compared to the semi-honest schemes.

\textbf{Implementation and Evaluation.}
We implement and evaluate Argos on real hardware\footnote{\url{https://github.com/mit-enclaves/argos}}.
Our prototype runs on Intel x86 platforms, but Argos is easily adaptable to other vendors and architectures (e.g., AMD X86, ARM or RISC-V) and compatible with most commodity machines past 2008.
To guarantee the TCB integrity at boot-time, we use commodity hardware root-of-trust technologies.
Our prototype uses a TPM, but our architecture is adaptable to other hardware roots of trust and secure coprocessors such as the Apple Secure Enclave~\cite{appleplatformsecurity} or Open Titan~\cite{johnson2018titan}.
Our security monitor is an open source fork of Tyche~\cite{castes2023creating}, modified to support our attestation scheme and remove side-channel protections, but our approach is also compatible with other micro-hypervisors like seL4~\cite{klein2009sel4}.
We implement a custom minimal runtime for FHE and the SEAL library~\cite{chen2017simple}, emulating system calls and providing randomness through a hardware random number generator (i.e., RDRAND instructions) that is not under OS control.
For more complex applications that require a broader class of system calls, Gramine~\cite{tsai2017graphene} can be used for better compatibility, at the price of a larger TCB and some performance overhead (5$\times$ slower startup time, for example).
Argos has a limited attack surface with a minimal trusted code base (TCB) of less than 18KLOC, plus the target FHE application ($\approx50$KLOC).
Our evaluation shows that Argos is 80 times faster than previous work leveraging Intel SGX for FHE integrity~\cite{viand2023verifiable} with a minimal average performance overhead of 3\% for FHE evaluation.
We show that Argos can be used to implement more complex protocols such as attested PIR and attested PSI with performance overheads under 8\% and without the significant offline communication costs incurred by cryptographic solutions.

\change{\textbf{Contribution.}
Argos is the first TEE-based platform to enable maliciously-secure verifiable FHE while being secure against all known microarchitectural side channels, all at minimal overheads.
Prior to this work, it was not obvious that TEEs in \emph{commodity hardware} could achieve such strong security guarantees.
Existing TEEs suffer from insecurity due to the fact that their remote attestation mechanisms are vulnerable to microarchitectural side channels (see Section \ref{sec:discussion}). 
As a result, TEEs have thus far remained undesirable for cryptographic applications such as verifiable FHE.
Argos is carefully designed such that only the TPM contains unencrypted secrets.
The TPM uses a simple microarchitecture and is microarchitecturally isolated from the CPU, blocking all known microarchitectural side channels. 
This means that we can leverage the TPM as a "signing oracle" to design a simple, efficient, and \emph{secure} remote attestation scheme.
Furthermore, this simple solution addresses an important problem: the efficient deployment of FHE in real-world scenarios with malicious security.}

\noindent{}To summarize, our main contributions are:
\begin{itemize}[leftmargin=*, align=left, itemsep=0ex, topsep=0ex]
\item 
    Argos, the first \emph{integrity-only} enclave platform designed to build maliciously-secure verifiable FHE; 
\item
    Argos can be used to build fully malicious and authenticated PSI and PIR schemes;
\item 
    Argos is secure by construction against microarchitectural side channels and transient execution attacks;
\item 
    Argos requires no specialized hardware and is compatible with commodity processors from 2008 onward; and
\item 
    Argos only incurs 3\% average performance overhead for FHE computation, less than 8\% performance overhead for complex protocols, and virtually no communication overhead;
\end{itemize}

%% file: sections/threat_model.tex
\section{Threat model}
\label{sec:threat_model}
Our goal is to ensure integrity of the data and computation running inside our enclave environment.
Most importantly, we do not protect confidentiality of our programs, nor do we protect them against denial of service.
We assume a strong adversary collocated on the server that has compromised the majority of the software stack including the OS, and can mount any microarchitectural side-channel and transient execution attacks.
Because Argos is secure against all software-mounted side channels that underlie known transient execution attacks, we will refer to all these attacks as ``microarchitectural side channels'' throughout the rest of the paper.
Our trusted computing base (TCB) consists only of a security monitor (18 KLOC) and our application, both assumed to be bug-free.
We assume that the hardware is \emph{functionally} correct and that the TPM's cryptography is properly implemented with constant-time programming.
Rowhammer and fault injection attacks are not as practical and are considered out of scope.
We also protect against the majority of physical attacks (cold boot, BIOS tampering, physical side channels) on the main processor and DRAM, but for the rest of this paper, we will focus on software-mounted attacks and consider physical attacks out of scope.
\change{A primer on microarchitectural and physical attacks can be found in Appendix~\ref{app:primerattacks}}.
A detailed discussion on the remaining attack surface can be found in \Cref{sec:discussion}.

%% file: sections/background.tex
\begin{figure*}[t]
\centering
\includegraphics[width=\textwidth]{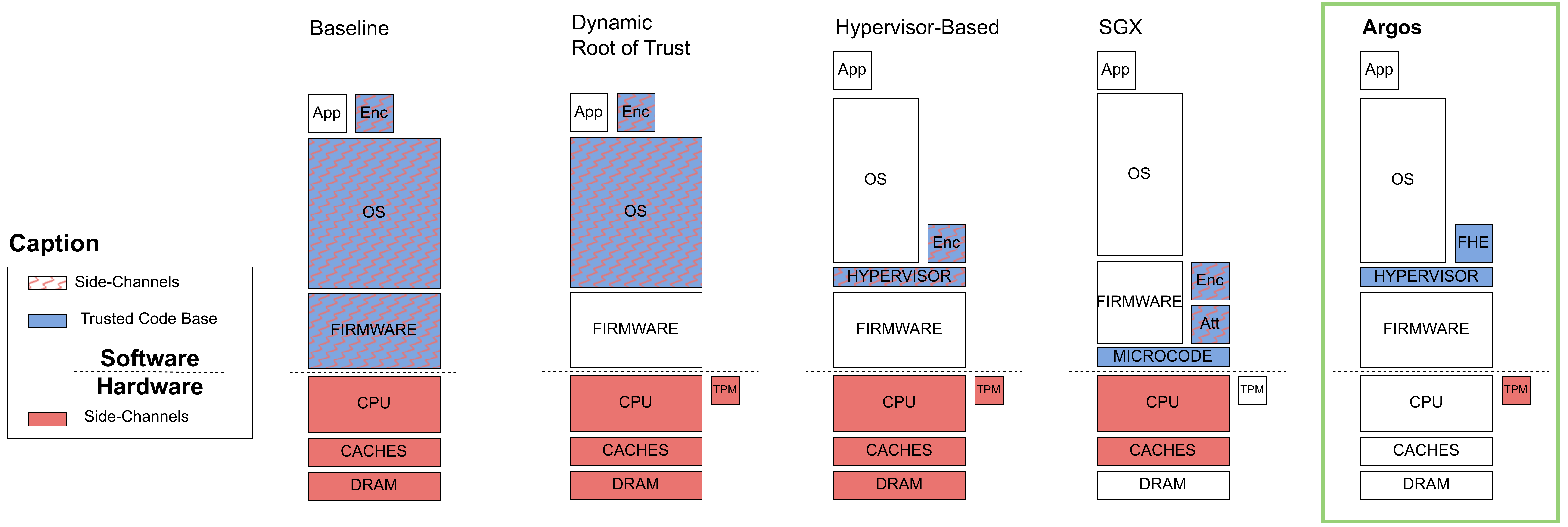}
\Description{Evolution of the attack surface on TEE platforms. Are presented: baseline, dynamic root of trust, hypervisor-based, SGX, and Argos. Argos has a significantly reduced attack surface.}
\caption{Evolution of the attack surface on TEE platforms. Enc: enclave program, Att: attestation enclave.}
\label{fig:computing_stack}
\end{figure*}

\section{Background \& Motivation}
\subsection{FHE Schemes And Client-Server Setup}
\label{sec:fhe_def}
We consider a client-server setup where a server evaluates a logical circuit on some client ciphertext obtained using an FHE scheme.
An FHE scheme is usually defined as a tuple of algorithms ($\Epsi.\text{Gen}$, $\Epsi.\text{Enc}$, $\Epsi.\text{Eval}$, $\Epsi.\text{Dec}$). The client generates a key pair using $\Epsi.\text{Gen}$ and encrypts some input $x$ using $\Epsi.\text{Enc}$.
The server then uses $\Epsi.\text{Eval}$ to evaluate a circuit $f$ on the ciphertext $c_x$ and sends the resulting ciphertext $c_y$ back to the client to decrypt using $\Epsi.\text{Dec}$.
Concretely:
\begin{itemize}
    \item $\Epsi.\text{Gen}(1^\lambda) \rightarrow (\text{pk}, \text{sk})$
    \item $\Epsi.\text{Enc}_{key}(x) \rightarrow c_x$
    \item $\Epsi.\text{Eval}_{pk}(c_x, f) \rightarrow c_y$ where $y=f(x)$
    \item $\Epsi.\text{Dec}_{sk}(c_y) \rightarrow y$
\end{itemize}
\noindent FHE schemes must enforce the following properties \change{(see Appendix~\ref{app:fheprop} for formal definitions)}:

\noindent\textbf{Correctness.} A scheme is \emph{correct} if any honest computation will decrypt to the expected result (i.e., ${Dec}_{sk}(c_y) = f(x)$ with high probability).

\noindent\textbf{Security.} All FHE schemes are secure against chosen plaintext attacks, which means that an attacker with the public key cannot distinguish between the encryption of two different messages of its choice.

\subsection{Semi-Honest vs. Malicious Security}
In a honest-but-curious or semi-honest setting, an attacker is allowed to observe intermediate messages and passively infer information in order to break security, but cannot deviate from the agreed-upon protocol.
In a malicious setting, the attacker is much more powerful and allowed to deviate arbitrarily from the protocol, e.g., to lie to the victim or tamper with messages.
The semi-honest setting is often used in theory, but has limited viability in a deployment context where assuming a malicious attacker is more realistic.

\subsection{Why Does FHE Need Verifiability?}
\label{sec:whyverifiablefhe}
Most state-of-the-art FHE schemes are \emph{insecure} to use in real-world settings~\cite{gentry2009fully, chillotti2020tfhe, ducas2015fhew, chen2017simple} \change{where malicious security is generally required}.
Because these schemes are not CCA-secure~\cite{bellare1998relations}, if an attacker is able to supply malformed ciphertexts and observe decryption outcomes (for example, by observing client behavior), it can mount key recovery attacks and completely compromise privacy~\cite{guo2024key, fauzi2022ind, chaturvedi2022practical, chenal2015key}.
Beyond the ``circuit-level'' security, the absence of integrity for FHE schemes also has implications for FHE-based applications.
A malicious server can arbitrarily modify the circuit they evaluate on the encrypted data.
Attacks on correctness are an obvious issue, but these can also translate to attacks on privacy at the application level.
Let us take the example of private contact discovery.
Alice wants to discover who among her contacts is using a private messaging service~\cite{signalcontact}.
Alice encrypts her address book, which contains all of her contacts' private information, and sends it to the server.
The server now has access to Alice's encrypted set of contacts.
If malicious, it could simply return the entire set to Alice.
This could mislead Alice into thinking that one of her friends (Bob) uses the messaging service.
Consequently, Alice contacts Bob via the service, leaking his phone number.
This would violate Bob's privacy and defeat the use of private contact discovery.

\subsection{Why Do We Need a New TEE Platform?}
Trusted execution environments (TEEs) provide hardware-based isolation for confidential programs while minimizing the trusted code base (see~\Cref{fig:computing_stack}). The existing TEE landscape spans from platforms that protect small ``enclave'' programs~\cite{AWSNitroEnclaves, pinto2019demystifying, costan2016intel} to those that implement confidential computing for entire virtual machines~\cite{li2022design, cheng2024intel, sev2020strengthening, ApplePrivateCloud}.
However, the discovery of numerous microarchitectural side channel and transient execution attacks has severely compromised their confidentiality guarantees.
Side channel attacks exploit shared microarchitectural structures such as memory caches \cite{gotzfried2017cache, brasser2017software}, translation look-aside buffers (TLB) \cite{gras2018translation}, branch predictors \cite{evtyushkin2018branchscope, aciiccmez2006predicting}, and DRAM controllers \cite{wang2014timing} to enable information leakage between security domains.

TEEs also provide integrity guarantees through remote attestation, allowing a remote client to verify the authenticity of the hardware and the initial state of the execution environment.
However, even these attestation mechanisms have proven vulnerable to side-channel attacks that can extract secret keys used by platforms to sign attestation reports~\cite{van2020sgaxe, chen2019sgxpectre, ryan2019hardware}, leading the cryptographic community to lose faith in TEEs.
On the other hand, discrete trusted platform modules (TPMs) and associated dynamic root of trust (DRoT) have long served as dedicated hardware for platform integrity and secure storage against software-based attackers.
However, their discrete nature also creates a severe performance bottleneck when attesting multiple security domains (Flicker~\cite{mccune2008flicker} incurs 2-3 orders of magnitude overhead). Although TPM virtualization~\cite{raj2016ftpm, AWSNitroTPM, AzureVirtualTPM, mccune2010trustvisor} can address these performance issues, in turn, they expose sensitive key material to microarchitectural side channels (see \Cref{sec:relatedwork} for a more detailed comparison of Argos and other platforms).

To rebuild trust in hardware security primitives, we must develop new systems that are both inherently secure against microarchitectural side channels and capable of outperforming functionally equivalent cryptographic solutions.
We address this challenge by focusing specifically on FHE applications and building a TEE platform that \emph{only enforces integrity} of the enclave program, which is sufficient for verifiable FHE since data remains end-to-end encrypted throughout the computation.

%% file: sections/overview.tex
\section{Insights \& Design Principles}

\subsection{FHE Applications Do Not Expose Secrets}
\label{sec:specificityfhe}

In FHE applications, no sensitive data is ever manipulated in clear.
This has several implications that can help simplify the design of a TEE platform and the attestation protocol.

\noindent\textbf{No Secrets Exposed.}
In a standard TEE platform, a central feature is to enforce the confidentiality of the execution environment.
In FHE, all data is encrypted.
As a result, the contents of our enclave program (i.e., the state of $\Epsi.\text{Eval}$) is public.
This means our application is not vulnerable to side channels, and we do not need to harden the execution environment against the usual threats.

\noindent\textbf{No Long-Term Secret Storage.}
In typical TEE platforms, an important feature is the ability for an enclave program  to \emph{seal} secrets and recover them later.
This property is usually tied to the identity (or the measurement) of the enclave program, and a secret can only be unsealed by a program that matches the correct measurement.
This is, for instance, what is used in Bitlocker to only ever release the disk encryption key to a correctly booted system.
TPMs and TEE platforms such as SGX all offer a seal operation.
FHE evaluation does not require long-term secret storage, which significantly reduces the features required by the TEE.

\noindent\change{\textbf{No Need To Pre-Establish Trust Before Sending Inputs.}}
In usual remote attestation protocols, the remote client first needs to verify the TEE attestation to establish trust.
Only once it trusts the TEE to have been correctly setup will it send its encrypted private input.
This is because the inputs will then be decrypted inside of the TEE.
In our case, inputs are never decrypted, which means that the remote client does not need to establish trust before sending the encrypted input.
This makes it possible to amortize the cost of attestation by only signing the final transcript of program inputs and outputs (see \Cref{sec:attesation}).

\subsection{Eliminating Microarchitectural Side Channels Using A Physical TPM}
Because our FHE-applications do not expose secrets, the only secret ever present on the server is the private signing key used in the attestation scheme.
This is a much smaller attack surface than in usual TEE systems.
We can take advantage of that opportunity to completely eliminate microarchitectural side channels by placing the signing key in a microarchitecturally-isolated physical TPM.
Because no secrets are left on the CPU or in main memory, this effectively makes our platform completely secure against microarchitectural side channels.

\noindent\textbf{Side Channels Are Read-Only.}
One important element to keep in mind is that side channels are ``read-only'' gadgets and can only extract secrets from a victim program.
Specifically, a side channel cannot modify the state of a victim's program memory or change its control flow.

\noindent\textbf{Side Channels Require Shared Resources.}
To mount a microarchitectural side channel, an attacker needs to trigger a transmitter that will access the secret and modulate a channel to transmit information to a receiver under the attacker's control.
That means the transmitter and receiver programs need to share some microarchitectural state in order for the transmission to be possible.

\noindent\textbf{Physical TPMs Are Microarchitecturally Isolated.}
TPMs are commonly implemented on a discrete chip or as firmware on a secure co-processor (we cannot secure software TPM running on the CPU).
They communicate with the main CPU using a hardware bus.
They do not share cache, nor any processor resources, with the CPU.
They use their own private resources, like private memory and registers, that are only accessible to the TPM.
As a result, if the secret key is kept in a physical TPM, it is microarchitecturally-isolated from the attacker and there is no possibility for the secret to be extracted through a CPU microarchitectural side channel.
\change{
In addition, the microarchitecture of TPMs and secure co-processors is intentionally kept simple, avoiding features like caches or out-of-order execution.
This prevents the existence of indirect microarchitectural side channels such as NetSpectre~\cite{schwarz2019netspectre}.}
Intrusive physical attacks might still be possible but are outside of our threat model (see ~\Cref{sec:discussion}).

\subsection{System Overview}
Our system is composed of several hardware and software elements (see \Cref{fig:computing_stack}).
The security monitor is a small ($\approx$18 KLOC) trusted piece of code running at the hypervisor privilege level.
It is in charge of enforcing isolation between the different security domains and integrity of the enclave programs.
Isolation is enforced by hardware mechanisms (see \Cref{sec:impl_isolation}) and integrity is enforced through our attestation scheme (\Cref{sec:attesation}).
The security monitor does not manipulate any secret key material and delegates all cryptographic operations that require a key to the physical TPM.
Integrity of the security monitor is enforced by the dynamic root-of-trust mechanism (\Cref{sec:impl_sm}).
Running on top of the security monitor, a large untrusted operating system is in charge of all resource allocations.
It contains all the complex logic, for instance, for memory allocation or scheduling.
It also provides many services for (enclave) applications such as networking, or I/O.

\noindent\textbf{Running an Enclave Program.}
To setup a trusted execution environment and securely run an enclave program, the untrusted OS
interacts with the security monitor through vmcalls to the hypervisor.
On behalf of the OS, the security monitor will reclaim some resources (memory and cores) and allocate them to the enclave.
Thanks to the isolation mechanisms under the control of the security monitor, the OS is now unable to modify the state of the enclave memory.
The security monitor will also enforce some invariants (setting the page table correctly and zeroing regions where no content is loaded), ensure the correct binary is loaded, and attest the enclave initial state (see \Cref{sec:detail_attest}).
Once loaded, the enclave can be started.
The remote client sends inputs over the network.
They are received by the OS and placed in shared memory so that the enclave can access them.
Inputs are systematically copied to enclave's private memory before being used.
Any output is also copied to shared memory and sent back to the client by the OS.

%% file: sections/remote_attesation.tex
\section{The Argos Attestation Scheme}
\label{sec:attesation}

\begin{figure*}[t]
\centering
\includegraphics[width=\textwidth]{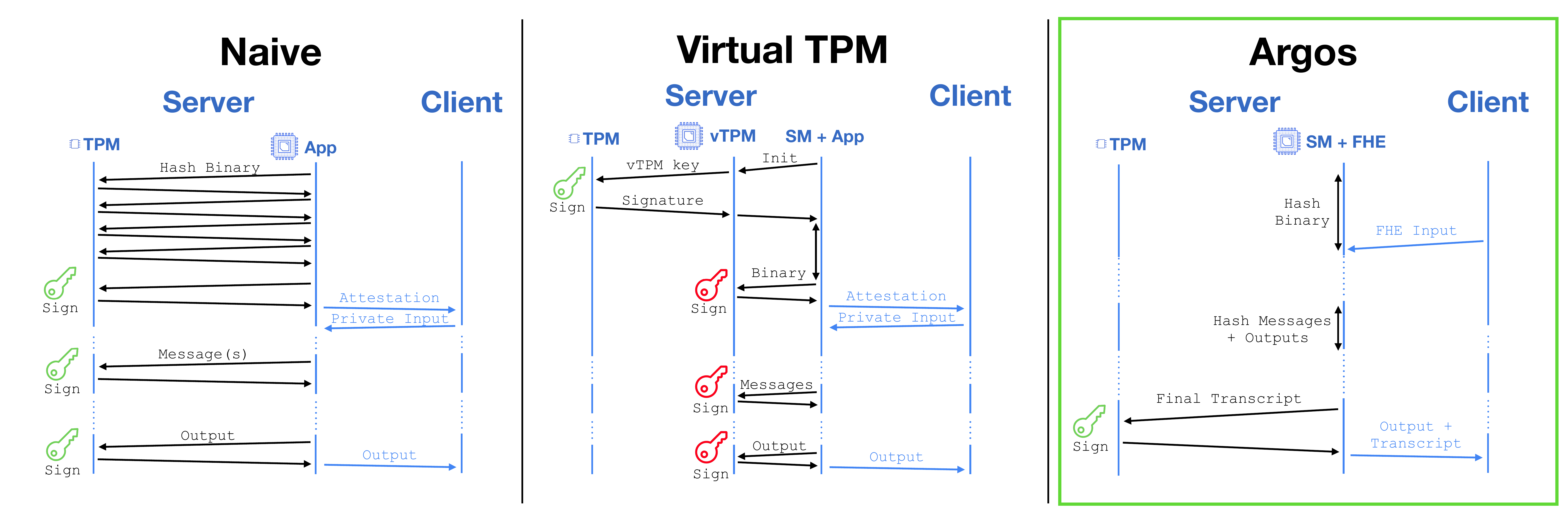}
\Description{Evolution of remote attestation protocols.}
\caption{\change{Evolution of remote attestation protocols. The naive approach with a secure co-processor is inefficient as it requires several back and forth communications with the discrete chip. The virtual TPM approach is insecure as it manipulates sensitive keys in the CPU, exposing them to microarchitectural side channels. The Argos protocol is both secure and efficient.}}
\label{fig:remoteatt}
\end{figure*}

\subsection{Existing Attestation Protocols}
Attestation schemes are generally interactive protocols that involve several rounds of communication between a client and a server.
First, before any sensitive data is sent, trust needs to be established between the execution environment running on the server side and the remote client.
The client will send a challenge and receive an attestation report, endorsed by the manufacturer and attesting that a genuine piece of hardware has been correctly set up to instantiate a trusted execution environment.
Once trust has been established, the enclave program and the remote client can perform key exchange and establish a trusted communication channel.
They can then start communicating and running the expected application.
Any message coming out of the enclave needs to be encrypted and attested using the private key generated inside of the enclave to prevent a malicious attacker from observing or tampering with the content of the messages.
Similarly, every message from the remote client needs to be decrypted and its integrity verified.

\subsection{Overhead of Using a Discrete TPM}
One of Argos's design principles dictates that no secret key material should ever be present on the main CPU.
This means that each enclave would need to delegate all key operations to a single physical TPM.
Even if only considering integrity protection (signature of messages and attestation protocol), this implies that the TPM becomes a serious bottleneck for the platform.
Performing a signature using a discrete TPM takes approximately 196 ms, and the numbers are similar for integrated TPMs.
This can potentially be a significant overhead for any application that requires several rounds of communication.
However, Argos's focus on FHE applications allows us to minimize much of this overhead.

\subsection{Attested Transcript for FHE Applications}
\label{sec:attestedtrans}
Trust does not need to be established before sending encrypted data.
Because intermediate results are also encrypted, privacy cannot be compromised until a ciphertext is decrypted.
That means integrity only needs to be checked \emph{just before} decrypting an output
(see \Cref{sec:circuit-levelvFHE} for our security analysis).
In other words, each message of the FHE application does not need to be individually attested.
Instead, the integrity-protected security monitor can keep a transcript of all relevant data (i.e., all application inputs, intermediate results, and final output), and, only perform a single TPM signature over that transcript.
This simplifies the attestation scheme and drastically reduces the cost of using a physical TPM as remote attestation is now a simple fixed cost \change{(see Figure~\ref{fig:remoteatt})}.

\subsection{Argos Attestation Scheme}
\label{sec:detail_attest}
\noindent\textbf{Measured Boot.}
Upon boot, the hardware dynamic root of trust (DRoT) is used to derive a measurement of the Argos security monitor (see \Cref{sec:impl_sm}).
This measurement is securely stored in the TPM registers (PCRs).

\noindent\textbf{Enclave Setup.}
When a new enclave is initialized, the security monitor will allocate some of its private memory to the enclave transcript.
During enclave setup, the security monitor measures the enclave binary and stores it in the transcript.
In our specific case, the binary contains the FHE circuit, small public inputs (e.g., FHE parameters), alongside the logic to evaluate the circuit (the SEAL library in our case).
It also contains the rest of the application logic.
All the values of the relevant architectural registers that represent the initial state of the execution environment are also added to this measurement.
For instance, the enclave initial program counter, page tables, the stack pointer and the addresses of shared memory used to communicate with the OS.
The measurement of the binary and initial state unambiguously represent a function that takes inputs from shared memory and outputs results to it.

\noindent\textbf{Enclave Lifetime.}
During the lifetime of the enclave, the transcript is appended with hashes of the data received and sent by the enclave.
This includes all inputs and outputs.

\noindent\textbf{Attesting the Transcript.}
Once the final output has been generated and the transcript extended with it, the application performs a special vmcall to the security monitor to request the attested transcript.
The security monitor then leverages the TPM to obtain a signature over the transcript.
The application will then be able to transmit the attested transcript to the remote client, along with the final output.

\noindent\textbf{Client Verification.}
Upon receiving the result(s) and the attested transcript, the client first verifies the attestation.
The verification of the attestation goes as follows:
First, the client verifies that the key used to sign the transcript belongs to a genuine TPM (see \Cref{app:primertpm} for more details).
It then checks that the measurement for the security monitor contained in the transcript signature matches a reference measurement that is known to be correct (see \Cref{sec:discussion} on how such a reference measurement can be obtained).
Finally, it checks that the binary reference in the transcript corresponds to the correct FHE application and that it was evaluated on the input it provided.
If all checks are satisfied, the verification is successful.
In case of a verification failure, the server is behaving maliciously and the client aborts.

\noindent\textbf{Output Decryption.}
If and only if verification succeeds, the client decrypts the FHE output.

\subsection{Remote Attestation as a Proof System}
Our simplified attestation scheme is now functionally equivalent to a proof system.
We can formalize it as a tuple ($\Pi.\text{Gen}$, $\Pi.\text{Prove}$, $\Pi.\text{Verify}$).
The server runs $\Pi.\text{Gen}$ to generate a key pair.
It can then use the private key to run $\Pi.\text{Prove}$ and generate an attested transcript ($\pi_y$) over some client input $x$, an application $g$ (formally a function, in practice a binary and an initial state for the execution environment), and the resulting output $y$.
Given the public key, the client can then run $\Pi.\text{Verify}$ to verify that the attested transcript $\pi_y$ is correct, that is, $y=g(x)$.
In summary, we have:
\begin{itemize}
    \item $\Pi.\text{Gen}(1^\lambda) \rightarrow (\text{pk}, \text{sk})$
    \item $\Pi.\text{Prove}_{\text{sk}}(x, g) \rightarrow (y, \pi_y)$ where $y=g(x)$
    \item $\Pi.\text{Verify}_{\text{pk}}(y, x, \pi_y) \rightarrow \texttt{true}/\texttt{false}$ where the client accept the output is \texttt{true}
\end{itemize}

\noindent Additionally, our scheme satisfies these two properties \change{(see Appendix~\ref{app:raprop} for formal definitions)}:

\noindent\textbf{Completeness.}
The system is \emph{complete} if $\Pi.\text{Verify}$ will always accept a honestly computed result, i.e., a correctly formed transcript where $y = g(x)$.

\noindent\textbf{Soundness.}
The system is \emph{sound} if an adversary cannot make $\Pi.\text{Verify}$ accept an incorrect answer i.e., a transcript where $y \neq g(x)$.
\\

\noindent\textbf{Modeling Several Rounds of Communication.}
For applications that require several rounds of communication, we only generate one attested transcript at the end (see \Cref{sec:attestedtrans}).
That means $x$ is the concatenation of all client inputs/messages.
Our specific implementation simply shares intermediate results (the server messages) with the client.
It also caches some internal state for $\Pi.\text{Prove}$ (such as the measurement of $g$) so it can efficiently generate the final attested transcript once provided with all client inputs.
\\

\noindent\textbf{Security Analysis.}
Completeness of our attestation scheme reduces to the \change{correctness} of the underlying signature scheme used in the TPM.
Soundness is more complex to establish, as it requires informally laying down all of our system security.
First, as we explained earlier, an attacker (as described in our threat model) is not able to extract the private key from the TPM.
Second, the security of our DRoT guarantees that the security monitor was correctly loaded and its integrity cannot be compromised.
Third, our assumption of the TCB tells us that the security monitor is correctly implemented and bug-free.
As a result, attested transcripts will only be generated for enclaves that are correctly set up, loaded, and executed.
Our assumption on the underlying hardware isolation mechanisms (execution modes and nested page tables) guarantees that the integrity of the enclave code and data is maintained over the course of its lifetime.
Finally, the functional correctness of the hardware guarantees that when an enclave is run that implements functionality $f$ on input $x$, the resulting output $y = f(x)$.
With all of this in place, the soundness of our attestation scheme reduces to the soundness of the TPM signature scheme.

%% file: sections/security_annalysis.tex
\section{Circuit-Level Verifiable FHE}
\label{sec:circuit-levelvFHE}
We first show how, using Argos, we can take a semi-honest FHE scheme and build a circuit-level verifiable FHE scheme that achieves malicious security.
In \Cref{sec:application}, we show how we can extend Argos to satisfy application-level security.

In our client-server setup, all state-of-the-art FHE schemes~\cite{chen2017simple, chillotti2020tfhe, ducas2015fhew} are \emph{insecure} when considering a malicious server.
Let us take a simple example where a client wants to outsource
some sensitive computation to a server using FHE.
Here, the server might be able to infer information regarding the decryption of a ciphertext by observing the client reaction.
For instance, in FHE, if the noise in the ciphertext overflows a given threshold, client-side decryption might fail, which could lead to the client emitting a new request to the server.
This is not a problem if assuming an honest-but-curious server.
However, if the server is malicious, it can tweak the response to the client in arbitrary ways and observe its reaction.
This is equivalent to providing our attacker with access to a (limited) decryption oracle that outputs ``success'' or ``failure'' for ciphertexts under the attacker's control.
This is already enough to mount key-recovery attacks~\cite{guo2024key, chenal2015key} and break security of the FHE scheme (if the attacker recovers the private key, it can now decrypt the client's private requests).
In fact, this is much more problematic than a simple ``correctness'' issue where the client would just get a wrong output.
In this simple outsourcing scenario, enforcing integrity at the \emph{FHE-circuit-level} is enough, as it prevents the attacker from providing malformed ciphertexts to be decoded.
When the client receives the response from the Argos server, it will first verify the attested transcript to ensure that the correct circuit was evaluated on the correct inputs.
If verification fails, the client aborts, does \emph{not} decrypt the ciphertext, and detects malicious behavior from the server.
If and only if verification succeeds, the client may decrypt the ciphertext with the guarantee that it was correctly formed.

\subsection{Definition}
\label{sec:defvfhe}
Inspired by the formalism of Viand et. al.~\cite{viand2023verifiable}, we define circuit-level verifiable FHE (vFHE) as follow.
A vFHE scheme is defined as a tuple of algorithms (Gen, Enc, Eval, Verify, Dec).
The client is equipped with a key pair $(\text{pk}_c, \text{sk}_c)$ and the server with $(\text{pk}_s, \text{sk}_s)$.
The client uses Enc to encrypt some input $x$ and obtains a ciphertext $c_x$.
Using Eval, the server evaluates a circuit $f$ over $c_x$ and generates an encrypted output $c_y$, along with an attested transcript $\pi_y$.
When the client receives the output, it will first \emph{verify} the transcript is correct ($c_y=f(c_x)$) and if and only if it is, use Dec to decrypt $c_y$ and recover $y$.
In summary:
\begin{itemize}
    \item $\text{Gen}(1^\lambda) \rightarrow ((\text{pk}_c, \text{sk}_c), (\text{pk}_s, \text{sk}_s))$
    \item $\text{Enc}_{\text{pk}_c}(x) \rightarrow c_x$
    \item $\text{Eval}_{\text{pk}_c, \text{sk}_s}(c_x, f) \rightarrow (c_y, \pi_y)$ where $y=f(x)$
    \item $\text{Verify}_{\text{pk}_s}(c_y, c_x, \pi_y) \rightarrow \rightarrow \texttt{true}/\texttt{false}$ where the client accept if \texttt{true}, aborts otherwise.
    \item $\text{Dec}_{\text{sk}_c}(c_y) \rightarrow y$
\end{itemize}

\change{Formal properties can be found in Appendix~\ref{app:vfheprop}.}
Correctness is defined identically to a semi-honest FHE scheme (see \Cref{sec:fhe_def}).
Security differs as our attacker is now malicious and can send arbitrary $c_y$ values.
We also add two properties compared to the vanilla FHE-scheme.

\noindent\textbf{Completeness.}
A vFHE scheme is \emph{complete} if Verify will always accept an honestly computed result.

\noindent\textbf{Soundness.}
Finally, a scheme is \emph{sound} if an adversary cannot make Verify accept an incorrect answer.

\subsection{Putting Everything Together}
Let's consider an FHE scheme ($\Epsi.\text{Gen}$, $\Epsi.\text{Enc}$, $\Epsi.\text{Eval}$, $\Epsi.\text{Dec}$) and our remote attestation scheme ($\Pi.\text{Gen}$, $\Pi.\text{Prove}$, $\Pi.\text{Verify}$).
Argos' construction is as follows:
\begin{itemize}
    \item $\text{Gen}(1^\lambda) \rightarrow ((\text{pk}_{\Epsi}, \text{sk}_{\Epsi}), (\text{pk}_{\Pi}, \text{sk}_{\Pi}))$ \\ with $(\text{pk}_{\Epsi}, \text{sk}_{\Epsi}) = \Epsi.\text{Gen}(1^\lambda)$ and $(\text{pk}_{\Pi}, \text{sk}_{\Pi}) = \Pi.\text{Gen}(1^\lambda)$
    \item $\text{Enc}(x) \rightarrow c_x$ with $c_x = \Epsi.\text{Enc}(x)$
    \item $\text{Eval}(c_x, f) \rightarrow (c_y, \pi_y)$ with ($c_y, \pi_y) = \Pi.\text{Prove}(c_x, \Epsi.\text{Eval}(~.~, f))$
    \item $\text{Verify}(c_y, c_x, \pi_y) \rightarrow b$ with $b = \Pi.\text{Verify}(c_y, c_x, \pi_y)$.
    \item $\text{Dec}(c_y) \rightarrow y$ with $y = \Epsi.\text{Dec}(c_y)$
\end{itemize}

\noindent The key here is to use the attestation mechanism to attest the correct execution of $\Epsi.\text{Eval}(~.~, f)$ represented by the binary and initial state of the execution environment.
\\

\noindent\textbf{Security Proof \change{--- Sketch}.}
Correctness of our construction reduces to the correctness of the initial FHE scheme.
Similarly, completeness and soundness directly derive from the completeness and soundness of the attestation scheme.
Security is more subtle.
Remember that our initial FHE scheme is not secure if provided with malformed ciphertexts.
That means we can only rely on the security of the underlying FHE scheme if our attacker cannot gain control over the ciphertexts.
This is enforced by the verification step.
If the client runs Dec on an output received from the server, that means the output and the transcript had to successfully pass Verify.
If this is true and the attacker was able to supply a ciphertext of its choice, i.e., $c_y\neq\Epsi.\text{Eval}(c_x, f)$, that means the attacker was able to break the soundness of the underlying attestation system, which is a contradiction.
As a result, our construction is secure, and we have successfully built a circuit-level verifiable FHE scheme.
\change{The full proof can be found in Appendix~\ref{app:securityfullcirtuitlevelfhe}.}
\\

\noindent\textbf{Batching FHE Evaluations.}
\label{sec:batching}
Running $\Pi.\text{Prove}$ is expensive (it requires a TPM sign operation).
Because evaluations of different FHE circuits are independent, we can batch them for performance optimization and only produce one attested transcript for all these evaluations.
Enc, Verify, and Dec are the same, and security still holds as long as Dec is never run if verification was not successful.

%% file: sections/applications.tex
\section{Extending Argos to FHE-Based Applications}
\label{sec:application}

Circuit-level security sometimes differs from application-level security.
For instance, attacks on correctness can escalate to attacks on privacy (\Cref{sec:whyverifiablefhe}).
Additionally, the security of our FHE schemes breaks if a malicious server is able to supply malformed ciphertexts for decryption (\Cref{sec:whyverifiablefhe}).
This is potentially dangerous when, in many FHE-applications, the server itself can provide public or private inputs and act as a party in the computation.
We explore two such applications, 1) private information retrieval, and 2) private set intersection.
For each application, we discuss the challenges in deploying FHE-based schemes in a real-world setting and show how Argos can be extended to enforce malicious security.

\subsection{Malicious Private Information Retrieval}
Private information retrieval (PIR) allows a client to retrieve an element from a public database hosted on a remote server, without the server learning which element was requested by the client.
PIR has many applications such as public-key directory~\cite{colombo2023authenticated}, certificate transparency logs~\cite{ryan2013enhanced,henzinger2023one}, private web search~\cite{henzinger2023private} or private analytics~\cite{green2016protocol}.
PIR can be instantiated in a multi-server or single-server setting using a range of cryptographic primitives, but in this paper we focus on the state-of-the-art for single-server PIR which are FHE-based schemes~\cite{angel2018pir, menon2022spiral, henzinger2023one, menon2024ypir}.

Most schemes assume a honest-but-curious server; however, this is not a realistic setup for real-world deployment.
Furthermore, circuit-level integrity alone is not enough to enforce malicious security.
A malicious server can always modify the values in the database, which could break not only the correctness (send the wrong answer for a given query), but most importantly security (i.e., privacy of client input) by mounting \emph{selective failure attacks}~\cite{kushilevitz1997replication}.
We show how Argos can be extended to secure semi-honest FHE-based PIR scheme in a malicious setting.

\smallskip
\vspace{0.05in}
\noindent\textbf{Defending Against Selective Failure Attacks.}
In these attacks, a malicious server chooses an entry in the database and replaces the entry with a value that it knows will cause a decryption failure.
This is possible because in FHE, using an input that is not in the correct plaintext space (for instance, an integer that is too large) can lead to a malformed ciphertext, which will not decrypt properly.
Because our attacker might be able to observe the client reaction to decrypting the output (e.g., the client will send the request again), if decryption fails, the attacker can infer with high probability that the client queried the malformed entry.

Circuit integrity is not enough here.
Argos also needs to preprocess the public database and check that the server's public inputs belong to the plaintext space.
If a check fails, Argos should abort the server execution and not serve any client request.
Formally, this is equivalent to allowing server inputs $w$ for $f$, and in our construction, wrapping $\Epsi.\text{Eval}$ in a function that evaluates $\Epsi.\text{Eval}$ if and only if $w$ is well formed and aborts otherwise.

\smallskip
\vspace{0.05in}
\noindent\textbf{Authenticated PIR.}
So far, our malicious PIR scheme is only able to guarantee that the server performed the protocol using a well-formed database, but not a specific one.
Adding authenticity to a PIR scheme is a desirable property and might be essential for some real-world deployments.
For instance, in the case of a transparency log or a public-key directory, it might be essential to guarantee the correct database was queried.
We take inspiration from previous work and solve this problem by having the server initially commit to the database. 
This assumes a reference commitment is available for the client to fetch out-of-band.\footnote{This assumption is made by Authenticated PIR~\cite{colombo2023authenticated} but recent works~\cite{dietz2024fully, de2024verisimplepir} have also shown how using a limited number of ``validation queries'', the client can verify with high probability that the committed database is correct.}
Argos can easily be extended to support this functionality by adding a fresh commitment to the database (i.e., a hash or the root of a Merkle tree) to the attestation transcript.
When receiving the response to its query, a client should first verify the attestation transcript, including that the database commitment matches the reference one.
If the verification fails, it should abort.
If (and only if) it succeeds, it should then decrypt the response and proceed.
In our formalism, this is equivalent to extending $\text{Verify}$ to take a reference commitment $H(w)$ as an input, and tweaking soundness to ensure an adversary cannot make Verify accept a transcript $\pi_y$ that does not contain the correct commitment.
\change{A detailed security \textit{argument} can be found in Appendix~\ref{app:secargpir}.}

\subsection{Malicious Private Set Intersection}
Private set intersection (PSI) allows two distrusting parties with some private sets to compute the intersection of these sets without learning more information than the intersection itself.
PSI has many applications, such as private contact discovery~\cite{signalcontact} or compromised credential checking~\cite{pal2022might}.

Most efficient PSI protocols do not achieve (full) malicious security~\cite{cong2021labeled, chen2017fast, chen2018labeled, kiss2017private}.
Some constructions do~\cite{nevo2021simple, rosulek2021compact,pinkas2020psi, de2010linear, raghuraman2022blazing, rindal2021vole, cryptoeprint:2024/570} but at the price of high offline communication cost (see \Cref{sec:evaluation_psi}).
Using a state-of-the-art semi-honest PSI~\cite{cong2021labeled} scheme based on FHE, we show how Argos can be extended to make the scheme fully secure against a malicious server.
Here, the main difference from the PIR setup is that the server inputs are not public, but private.
State-of-the-art FHE schemes are not \emph{circuit private}. 
That means that the client might be able to learn information regarding the server's private input by looking at the noise contained in the output.
FHE-based PSI schemes mask the server's input using oblivious pseudo random functions (OPRFs) to provide privacy for the server even in the presence of malicious clients.
Moreover, the security of the FHE scheme is argued to also provide privacy for the client, even in the presence of a ``malicious'' server.
This is because the ideal PSI functionality considered in these works assumes a limited attacker with no observability of the client reactions~\cite{chen2018labeled}.
However, this might not hold in a real-world deployment, as we explained before.

\vspace{0.1in}
\noindent\textbf{Verifiable PSI.}
Once deployed in a real-world setting, the attacker might be able to observe some of the client's reactions.
Attacks on correctness might have serious security implications, sometimes translating to attacks on privacy (see example of private contact discovery in \Cref{sec:whyverifiablefhe}).
Fortunately, circuit-level integrity---as provided by Argos---can guarantee \emph{verifiable} PSI, which means that verification of the attestation transcript succeeds if and only if the server correctly executed the PSI circuit on a private set.
Nevertheless, for full malicious security, we still need to ensure that the server evaluates PSI over the \emph{correct} set.

\vspace{0.1in}
\noindent
\textbf{Authenticated PSI.}
If the server's input set is not authenticated, a malicious server can arbitrarily modify the dataset without detection.
This can have serious security implications in deployment.
For instance, in the case of compromised credential checking, Alice reaches out to Mozilla's server to ensure that her passwords were not compromised and are safe to use.
If Alice indeed uses some compromised credential and the server is malicious, it could evaluate the PSI circuit on her private set, but using a malicious input set, crafted server side, ensuring no intersection is found.
This might lead Alice to use her compromised credentials.

Similarly to Authenticated PIR, Argos can be extended to include a commitment to the dataset (e.g., a Merkle tree or hash-based commitment scheme) in the attestation transcript.
Here too, a public commitment needs to be made available out-of-band to the client, through a transparency log, for instance.
Upon receiving a response to its request, the client first verifies the transcript and aborts if it does not match the correct commitment.
Formally, the extensions we made for PIR are almost sufficient, we just need to tweak the wrapper function over $\Epsi.\text{Eval}$ to match our PSI functionality.
Instead of only checking that $w$ is well formed, the wrapping function should compute the corresponding OPRF values for $w$.
The server's privacy is guaranteed by the OPRF, the soundness of the attestation scheme, and the fact that commitment schemes are hiding i.e., they do not leak information about the server's set.
\change{The security \textit{argument} can be found in Appendix~\ref{app:secargpsi}.}

%% file: sections/implementation.tex
\section{Implementation Details}
\subsection{Security Monitor}
\label{sec:impl_sm}
Our security monitor is based on the X86 version of Tyche~\cite{castes2023creating}, a Rust micro-hypervisor that exposes hardware resources and isolation primitives to better protect trust domains.
Note that other ``small'' hypervisors such as XEN~\cite{barham2003xen} are at least one order of magnitude larger or closed source.
We implement our attestation scheme on top of Tyche, including support for the hardware TPM and the interface to manage attested transcripts.
We are also able to simplify some aspects of the platform, for instance, by removing side channel protections from Tyche (e.g., cache and CPU flush).
One of the implementation challenges was to adapt Tyche and the Linux driver to support large (i.e., several GBs) enclaves with our custom runtime and manage memory allocation accordingly.   
To enforce the integrity of the security monitor, we use the TPM and a dynamic root of trust (e.g., Intel TXT or AMD DRMT~\cite{futral2013intel, nie2007dynamic}) to implement standard measure boot (see primer in \Cref{app:primertpm}).
At the end of this process, two TPM registers (PCRs) are set with hash values that uniquely identify the security monitor binary.
Later, when the security monitor needs to attest an application transcript, it will load the transcript hash in a third PCR and request a TPM Quote (a signature over select PCRs).

\subsection{Memory Isolation Primitives}
\label{sec:impl_isolation}
We use hardware extensions for virtualization to enforce isolation between security domains~\cite{van2006hardware}.
Extended or nested page tables~\cite{bhargava2008accelerating}, as implemented by Intel VT-x or AMD-V, make it possible for a guest domain to have an entire address space available and create a level of indirection for memory management that is exclusively under the control of the hypervisor.
That means no execution domain can access other execution domains' address space nor modify its content.
Other virtualization technologies such as AMD-Vi and Intel VT-d make it possible to protect domains from arbitrary DMA accesses.
We use these technologies in Argos to enforce memory isolation for our enclave programs.

\subsection{Custom Runtime}
\label{sec:impl_runtime}
The execution environments provided by Argos are similar to bare-metal environments with no OS support.
That means no syscalls, no program loader, no memory allocation, nor access to shared libraries.
A runtime is required to support programs and modern libraries like SEAL that leverage many services from the OS.
Gramine \cite{tsai2017graphene} is such a runtime and is already supported by Tyche.
It was conceived to securely run off-the-shelf applications on Intel SGX.
As a result, it implements all Linux syscalls and also has extensive support for shared library and memory management.
However, because of its versatility, it is rather large (20KLOC) and adds some non-negligible overhead~\cite{GraminePerfromance}.
In order to reduce the TCB and improve performance, we implement our own custom runtime. 
It is minimal (870LOC) and consists of a simple memory allocator~\cite{peterson1977buddy} and a handful of syscall handlers required by our application.
For instance, we implement our own \texttt{open} and \texttt{read} syscalls to supply randomness from \texttt{RDRAND} when an application attempts to read \texttt{/dev/urandom}.
We write our runtime to interface with MUSL~\cite{MUSL} by interposing on system calls and providing our own handlers.
That means all applications and libraries (including SEAL~\cite{chen2017simple}) need to be compiled statically and linked using our MUSL library, which requires some engineering effort to port applications but makes it possible to obtain high performance \emph{and} a very light runtime.

%% file: sections/evaluation.tex
\section{Evaluation}

\subsection{Testbench}
We prototype Argos on a \local{local Dell Optiplex machine} from 2017 equipped with an Intel Core i7-7700 processor clocked at 3.6GHz, a discrete TPM 2.0, and 8GB of RAM.
We run our security monitor (adapted from Tyche~\cite{castes2023creating} see \Cref{sec:impl_sm}) directly on the hardware and evaluate our benchmarks using our custom SEAL runtime or Gramine~\cite{tsai2017graphene}.
We run Ubuntu 22.04 with the Linux kernel v6.2 and SEAL 4.1 with the BFV scheme~\cite{brakerski2014leveled}.
All benchmarks are single-threaded, compiled using clang++ with the O3 optimization level.

We compare Argos with two other TEE platforms that have the following specs.
SGXv2 running on an \azure{Azure server} equipped with an Intel Xeon Platinum 8370C CPU with a frequency of 2.80GHz.
AWS Nitro Enclaves~\cite{AWSNitroEnclaves} running on an \aws{AWS server} equipped with an Intel Xeon Platinum 8259CL CPU with a frequency of 2.50GHz.
The SGXv2 setup uses Gramine as a runtime while Nitro Enclave uses a docker container environment that runs its own Linux kernel.
Note that under our threat model, both SGXv2 and AWS Nitro Enclaves are \emph{insecure} and are only used as comparison points for performance.
See \Cref{tab:tee_platforms} for a more complete comparison.
All numbers are obtained by averaging values over 10 runs.

\begin{table}
\begin{minipage}[t]{0.90\linewidth}
\centering
\begin{tabular}{|p{4cm}|r|}
\hline
Component & LOC\\
\hline\hline
\rowcolor{LightRed}
BIOS & 1.5M\\
\hline
\rowcolor{LightRed}
Linux & 28M\\
\hline
Security Monitor & 18K\\
\hline
Runtime (Custom / Gramine) & 870 / 20K \\
\hline
SEAL Library~\cite{chen2017simple} & 20K \\
\hline
Application & 1K---20K \\
\hline
\end{tabular}
\caption{Software Components and TCB Breakdown.
Red columns are excluded from the TCB.}
\label{tab:tcbloc}
\end{minipage}
\centering
\begin{minipage}[t]{0.35\linewidth}
\centering
\begin{tabular}{|c|c|} \hline
\textbf{Platform} & \textbf{Time}\\ \hline
Software & 42\\ \hline
vTPM & 136\\ \hline
dTPM & 195752\\ \hline
\end{tabular}
\caption{Signing operation ($\mu$s).}
\label{tab:tpm_signing}
\end{minipage}
\hfill
\begin{minipage}[t]{0.60\linewidth}
\centering
\begin{tabular}{|c|c|c|c|} \hline
& \textbf{Setup} & \textbf{Attest} & \textbf{Term}\\ \hline
\azure{SGX2} & 953 & 7 & 274\\ \hline
\aws{Nitro} & 3759 & 1 & 363\\ \hline
\local{Argos+G} & 466 & 196 & 5\\ \hline
\local{Argos} & 85 & 196 & 10\\ \hline
\end{tabular}
\caption{Fixed-cost operations on a 1GB enclave (ms).}
\label{tab:enclave_setup}
\end{minipage}
\end{table}

\subsection{TCB Evaluation}
Detailed count for lines of code (LOCs) can be found in Table~\ref{tab:tcbloc}.
Our TCB is small with a minimum of 40KLOC for our PIR application and a total of 60KLOC for our PSI application.
Note that we did not account for \change{SMM code,} the microcode used in the DRoT, or the TPM as it is not public information, and we consider these elements as part of the hardware.
This should be compared to other TEEs and system projects that try to keep a small footprint.
For instance, Nitro Enclaves include an entire Linux kernel which blows up their TCB in the range of MLOC, several orders of magnitude bigger than our TCB.
On the other hand, SGX would not include the \change{SMM code or the }security monitor (18KLOC), but relies on a significant amount of microcode.
As a point of comparison, XEN~\cite{barham2003xen} (a popular mini-hypervisor), and Coreboot~\cite{borisov2009coreboot} (a minimal BIOS) each total 200KLOC.
\change{On the other hand, micro-kernels such as SeL4~\cite{klein2009sel4} are in a similar size range with tens of thousands of lines of code.}

\subsection{Microbenchmarks}
\label{sec:microbenchmarks}

\noindent\textbf{TPM Sign Operation.}
\label{sec:eval_tpm}
We evaluate the cost of signing a 32-byte message (i.e., a hash) using a physical TPM.
We also compare the incurred overhead with two insecure baselines: 1) software signature in the application and 2) software signature performed by the security monitor (vTPM).
Results can be found in Table~\ref{tab:tpm_signing}.
The vTPM signature requires a vmcall and a context switch to the security monitor, which incurs a small overhead (about 100$\mu$s) compared to the application-level signature.
Operations with the physical TPMs are comparatively slow: we measure 196ms to perform a signature (which in our case boils down to loading the hash value in a PCR and requesting a TPM quote).
This is 3 orders of magnitude slower than a vTPM.
Some modern chipset-integrated hardware TPMs are faster than discrete TPMs and are still microarchitecturally isolated from the CPU.
However, they are only 5$\times$ faster than the discrete TPM at best.
These results hint at the importance of Argos' optimized attestation scheme, which only requires at most one signature per FHE evaluation and even less when using batching.
\\

\noindent\textbf{Enclave Fixed Costs.}
We evaluate the different fixed costs of different operations over the lifetime of an enclave.
To obtain measurements, we instrument a dummy enclave for which we allocate 1GB of memory.
Setup time is measured between the time a command line program requests the operating system to start an enclave and when execution reaches the main function of the enclave application.
This means it includes the allocation of enclave resources, loading of the binary, page table setup, binary measurement, and runtime initialization.
We can see that Nitro Enclaves have the highest setup time mostly due to the fact that they contain an entire Linux kernel that needs to be booted.
SGXv2 also presents some large startup time.
This is mostly due to all the special instructions required to start an enclave and the time it takes Gramine to initialize.
As indicated in Gramine's documentation, initialization can initiate more than 200 OCALLs, which all incur expensive context switches with architectural flushes (CPU caches and state) that cost 8,000 --- 12,000 cycles each~\cite{GraminePerfromance}.
This overhead can also be observed when running Argos with Gramine.
In comparison, our custom runtime is initialized with minimal overhead, which might be essential for latency-sensitive applications.

Attestation time is the time it takes for an enclave to request an attestation quote.
For SGXv2, support for remote attestation service by Intel is deprecated~\cite{IntelSGXEPIDDeprecated} so we use the analogue Microsoft Azure Attestation~\cite{GramineMAA} scheme.
Once provisioned, signing of the attestation quote is performed locally by the quoting enclave, hence the high performance similar to using a vTPM.
For Nitro Enclaves, the attestation is performed by the hypervisor using a vTPM~\cite{AWSNitroTPM} which also explains the high performance.
Due to Argos's reliance on a physical TPM, we have a comparably slow attestation time.

Termination time is measured between when the enclave main function exits and when the control returns to the command-line program that launched the enclave.
It includes all the time required to deallocate, clear resources, and return them to the OS.
For SGX, that also implies a lot of cleanup at the hardware level.
For Nitro Enclaves, it requires to wait for Linux to shutdown, which also adds delay.
In Argos, since all sensitive data is encrypted, memory does not need to be sanitized before being reallocated to the OS.
\\

\noindent\textbf{Transcript Size.}
Our transcript is quite small and composed of 2 PCR values used by the DRoT, the initial value of some key registers for the execution environment, the hash of our binary, the hash of messages exchanged by the application, the hash of some server input if needed.
It also contains the signature and the TPM public key along with the TPM certificates endorsed by the manufacturer.
Concretely, we measure a small transcript size of only 1,407 bits, negligible in front of the size of an FHE ciphertext ($\approx$500KB).

\subsection{FHE Evaluation}
\begin{table}
\small
\centering
\begin{tabular}{|l|c|c|c|} \hline
\textbf{Platform}&  \textbf{Tiny}&  \textbf{Small}& \textbf{Medium}\\ \hline
\hline
\viando{Baseline~\cite{viand2023verifiable}}&  \old{2ms}&  \old{11ms}& \old{14ms}\\ \hline
\viando{Bulletproofs~\cite{bunz2018bulletproofs}}&  \old{7569s}&  \old{3957s}& \old{8697s}\\ \hline
\viando{Aurora~\cite{ben2019aurora}}&  \old{1554s}&  \old{3750s}& \old{5028s}\\ \hline
\viando{Groth16~\cite{groth2016size}}&  \old{196s}&  \old{473s}& \old{634s}\\ \hline
\viando{Rinocchio~\cite{ganesh2023rinocchio}}& \old{320ms}& \old{305s}&\old{443s}\\ \hline
\viando{SGXv1~\cite{viand2023verifiable}}& \old{154ms}& \old{1100ms}& \old{1260ms}\\ \hline
\hline
\azure{Baseline Azure}&  283us&  1727us& 3170us\\ \hline
\azure{SGXv2 Azure}&290us (\azure{+3\%})&1840us (\azure{+7\%})&3638us (\azure{+15\%})\\ \hline
\hline
\aws{Baseline AWS}&  324us&  1889us& 3456us\\ \hline
\aws{Nitro Enclave}& 317us (\aws{-2\%})& 1827us (\aws{-3\%})& 3450us (\aws{-0\%})\\ \hline
\hline
\local{Baseline Local}& 351us& 2341us&4376us\\ \hline
\local{Argos+G}& 392us (\local{+12\%})& 2702us (\local{+16\%})&5202us (\local{+19\%})\\ \hline
\local{Argos}& 352us (\local{+0\%})& 2480us (\local{+6\%})&4447us (\local{+2\%})\\ \hline
\end{tabular}
\caption{Time for FHE evaluation for three circuits. \change{Baseline is always semi-honest FHE \textit{without} integrity.} Greyed values are reported from~\cite{viand2023verifiable}.
We indicate the overhead compared to the baseline on the same machine.} 
\label{tab:microbenchmarkl}
    \centering
    \begin{tabular}{c|>{\centering\arraybackslash}p{0.8cm}|c|c|c|c|c|}
            \cline{2-7}
          & \multicolumn{3}{c|}{Per Enclave} & \multicolumn{2}{c|}{Per Query} & \multicolumn{1}{c|}{Per Batch} \\ \cline{2-7}
         & Setup & Pre-Proc.& Hash & Proc. & Hash & Attest. \\ %\cline{2-6}
         \hline
         \multicolumn{1}{|c|}{Baseline} & 3 & 2835 & N/A& 1593 & N/A & N/A \\ \hline
         \multicolumn{1}{|c|}{Argos+G} & 1559 & 3192 & 465& 1812 & 1 & 196 \\ \hline
         \multicolumn{1}{|c|}{Argos} & 158 & 2821 & 465& 1610 & 1 & 196 \\ \hline
    \end{tabular}
    \caption{Breakdown of end-to-end execution times for authenticated private information retrieval (ms).}
    \label{tab:results_pir}
    \centering
    \begin{tabular}{@{}l@{}c@{}|c|c|c|c|c|c@{}|}
        \cline{3-8}
          &&  \multicolumn{3}{c|}{Enclave}&\multicolumn{2}{c|}{Query}& \multicolumn{1}{c|}{Batch} \\ \cline{2-8} 
  &\multicolumn{1}{|c|}{\#C}& Setup &Pre-Proc.&Hash&Proc.&Hash&Attest.\\ %\cline{2-7}
  \hline 
         \multicolumn{1}{|c}{Baseline} &\multicolumn{1}{|c|}{1}& 4&40351&N/A&3231&N/A&N/A\\ \hline  
         \multicolumn{1}{|c}{Argos+G} &\multicolumn{1}{|c|}{1}& 1576& 44530& 132& 3732& 187& 196\\ \hline
         \multicolumn{1}{|c}{Argos} &\multicolumn{1}{|c|}{1}& 179& 37054& 105&3354& 148& 196\\ \hline \hline
         \multicolumn{1}{|c}{Baseline} &\multicolumn{1}{|c|}{3K}& 4&40348& N/A&3276&N/A&N/A\\ \hline  
         \multicolumn{1}{|c}{Argos+G} &\multicolumn{1}{|c|}{3K}& 1530& 44467& 131& 3740& 187& 196\\ \hline
         \multicolumn{1}{|c}{Argos} &\multicolumn{1}{|c|}{3K}& 175& 36933& 104& 3403& 148& 196\\ \hline
    \end{tabular}
    \caption{Breakdown of end-to-end execution times for private set intersection (ms). \#C: size of the client set.}
    \label{tab:result_psi}
\end{table}

We evaluate the overhead of circuit-level vFHE in Argos compared to semi-honest FHE.
Results can be seen in \Cref{tab:microbenchmarkl}.
We use the same benchmarks as previous work on vFHE~\cite{viand2023verifiable} and compile and evaluate the different circuits using the SEAL library~\cite{chen2017simple}.
The results from Viand. et al. highlight the impractical performance overheads when using cryptographic proofs \change{such as Bulletproofs~\cite{bunz2018bulletproofs}, Aurora~\cite{ben2019aurora}, Groth16~\cite{groth2016size} or Rinocchio~\cite{ganesh2023rinocchio},} to build vFHE: between 6 and 7 orders of magnitude, with the exception of \change{Rinocchio} for the Tiny benchmark.
Their \change{semi-honest} baseline is also relatively slow, hinting that they might have disabled hardware acceleration.
Their implementation of FHE-in-TEE also shows poor performance probably due to the limitation of SGXv1 (limited 128MB EPC memory and unreliable timers).
For a fair comparison, we run the benchmarks of SGXv2 and Nitro Enclaves.
We compiled the benchmarks on one machine and used the same binary across or different setups.
Here, the performance looks much better with single-digit overheads and even small speedups for Nitro Enclaves.
Indeed, once setup, running a program in an enclave is equivalent to running it with bare-metal performance.
We tried to keep the runtime as similar as possible between the different configurations, but small differences might explain these results.
For Argos, we evaluate two different configurations. One using the Gramine runtime (Argos+G) and one using our custom runtime.
Our custom runtime performs better than Gramine.
This might be for a couple reasons including that the MUSL library tends to be faster than Glibc, our custom runtime simply ignores interrupts sent by the OS scheduler (the enclave is not de-scheduled), and our custom syscall handlers handle less corner cases and are more efficient (for instance, to allocate memory).
In conclusion, Argos shows performance comparable to commercial TEEs while offering better security.
Our custom runtime makes it possible to reduce the TCB and improve performance at the cost of some engineering effort to port applications.

\subsection{Authenticated Private Information Retrieval}
\label{sec:evaluation_psi}
We evaluate the overhead of authenticated PIR using Argos compared to semi-honest unauthenticated PSI.
We use the SealPIR library from Angel et.~al.~\cite{angel2018pir}.
We use the BFV scheme with polynomials of degree 4096. The plaintext modulus is set to 20. 
We evaluate performance for a database of size $2^{20}\approx1\text{M}$ where each element measures 128 bytes.
We allocate 4GB of static memory to our enclave.
We instrument the code to hash the database, queries, and responses and to attest the final transcript once all queries have been processed.
Results can be found in \Cref{tab:results_pir}.
We compare two configurations for Argos.
One using the Gramine runtime (Argos+G) and one using our custom runtime.
For Argos+G, the server setup  --which is a fixed cost-- shows an overhead of $+83\%$ mostly due to overhead setting up the enclave (+55\%) and hashing the database (+16\%).
For query processing, we measure an overhead of $13\%$ for fetching one element.
Here, hashing the response before sending it to the client takes a negligible amount of time.
Taking into account the signing of the final transcript, the overhead goes up to $26\%$ for retrieving one element.
Batching queries here would amortize costs down to a minimum of $13\%$ for numerous queries.
For Argos using our custom runtime, overhead goes down to 21\% for setup (mostly due to hashing of the batabase) and 1\% for query processing.
This is due to our more efficient handling of system calls.
The only cost left is attestation that can be amortized by batching requests (see \Cref{sec:batching}).
Once again, our custom runtime offers better performance than Gramine.
For all setups, Argos's transcript incurs negligible (>1\%) communication overhead.

Recent work on single-server authenticated PIR using cryptographic constructions by Colombo. et al., while providing better security guarantees, show a performance overhead of 30-100$\times$~\cite{colombo2023authenticated}.
Other recent work, VeriSimplePIR~\cite{de2024verisimplepir}, shows more reasonable online overheads (13\%-20\%) but incurs significant offline communication of the same order of magnitude as the database itself (2GiB communication for a 4GiB dataset). 
This also makes updating the database extremely difficult, while our scheme only requires updating a Merkle tree and communicating the commitment.

\subsection{Authenticated Private Set Intersection}
We evaluate the overhead of authenticated PSI using Argos compared to semi-honest non-attested PSI.
We use the APSI library compatible with SEAL that implements the state-of-the-art FHE-based scheme from Cong et.~al.~\cite{cong2021labeled}.
We use the BFV scheme with parameters 8192 for the polynomial degree modulus and the coefficient modulus. The plaintext modulus is set to 65537. 
We evaluated the performance of the unlabeled scheme with a server set size of $2^{20}\approx1M$.
We allocate 4GB of static memory to our enclave.
We instrument the code to hash the database, queries, and responses and to attest the final transcript once all queries have been processed.
Results can be found in \Cref{tab:result_psi}.
For Argos+Gramine, the server setup phase, we measure an overhead of $15\%$ mainly due to preprocessing operations, but with our custom runtime, we see a \emph{speedup} of 8\% due to our fast handling of memory allocation, and other syscalls.
Hashing the database is negligible here.
For query time, we see +16\% and +8\% overhead for Gramine and our custom runtime, respectively.
For all setups, Argos incurs negligible communication (>1\%).
In comparison, the state-of-the-art for malicious PSI using a cryptographic construction~\cite{raghuraman2022blazing} shows a 19\% performance overhead and a 67\% communication overhead compared to their own semi-honest construction for a dataset of $2^{20}$ elements. However, they cannot guarantee that the server uses the same dataset across clients and require to compute and transfer a large commitment for each client.

\noindent\textbf{Attested Transcript.}
Our remote attestation scheme makes it possible to amortize the cost of attestation over messages and queries.
In the PSI scheme we use, each query requires two rounds of communication (one round for OPRF, and one for the query itself), plus one initial attestation, which would bring the overhead for one query to $+27\%$.
Instead, we only need to pay the cost of one attestation operation per batch.
The PSI scheme we use already supports batching, but only up to a certain client set size (this is a limitation of FHE-based PSI schemes).
For our given server set size, we empirically measure a non-negligible failure rate for client set size above $3000$ elements.
A client who would want to compute the intersection of a bigger set would have to do so using several queries. 
That means that the maximum overhead of $15\%$ for a client set size of 1 can be amortized for larger client set sizes.
For instance, for a client set size of $30,000$, computing the complete intersection would take 10 queries and overhead would be reduced back to $8\%$.
This phenomenon would be heightened for more efficient PSI schemes as the cost of TPM signing is constant.

\subsection{Analysis of the Remaining Attack Surface}
\label{sec:discussion}
Argos primarily defends against microarchitectural side channels, which constitute the vast majority of TEE attacks so far (43 published according to \cite{li2024sok}).
While Argos is vulnerable to fault injection \cite{qiu2019voltjockey, murdock2020plundervolt, kenjar2020v0ltpwn} and Rowhammer attacks (less studied on TEEs), these threats are not fatal -- software can be hardened against fault injections~\cite{breier2022practical}, and no Rowhammer attacks have demonstrated the capabilities needed to compromise Argos (gaining hypervisor privilege on a hardware VM).
Regarding physical attacks, Argos protects against ColdBoot \cite{halderman2009lest} and physical side channels on the CPU~\cite{lee2020off}, but remains vulnerable to physical fault injection attacks \cite{chen2021voltpillager, buhren2021one}.
\change{The TPM only exposes a minimal attack surface for physical side channels, and modern TPMs are now directly integrated into the SoC, significantly raising the bar for these attacks~\cite{jacob2023faultpm}.
Indeed, integrated TPMs do not implement modern features such as DVFS or RAPL, making all known software attacks such as Hertzbleed~\cite{wang2022hertzbleed} or PLATYPUS~\cite{Lipp2021Platypus} and others~\cite{zhang2021red} impossible.
Other physical attacks are out-of-scope, but would still be significantly less practical:
1) CPU-noise most likely masks signals from small integrated TPM, 
2) TPM cryptography is assumed \emph{constant-time}, thwarting simple power and EM attacks~\cite{mantel2018secure, genkin2015get},
3) for \emph{constant-time} code, side channels are sometimes due to advanced circuit optimizations~\cite{lipp2022amd}, unlikely in a small integrated co-processor.}
Finally, our TCB might still contain bugs, and attacks that exploit implementation mistakes have been demonstrated on the Intel DRoT and the associated secure coprocessor~\cite{wojtczuk2009attacking, wojtczuk2009another, ermolov2017hack, IntelCSMEQSRAdvisory, buhren2019insecure}.
\change{However, these implementation errors can be \textit{patched when discovered} ---unlike vulnerabilities to microarchitectural side channel attacks.}

%% file: sections/relatedwork.tex
\begin{table*}
    \begin{minipage}[t]{1.0\textwidth}
    \centering
    \small
    \begin{tabular}{|>{\centering\arraybackslash}m{2.2cm}|*{6}{c|}*{3}{c|}c|c|c|}
    \hline
    \multirow{2}{*}[-4ex]{\begin{tabular}[c]{@{}c@{}}\textbf{TEE Platform}\end{tabular}} & 
    \multicolumn{6}{c|}{\textbf{Security}} & \multicolumn{3}{c|}{\textbf{Usability}} & \multicolumn{3}{c|}{\textbf{Performance}} \\
    \cline{2-13}
    & TCB &
    \rotatebox{90}{\parbox[c]{0.8cm}{\centering $\mu$-arch SC}} &
    \rotatebox{90}{\parbox[c]{0.8cm}{\centering Cold Boot}} &
    \rotatebox{90}{\parbox[c]{0.9cm}{\centering Phys. \\ SC CPU}} &
    \rotatebox{90}{\parbox[c]{0.8cm}{\centering Fault Inject.}} &
    \rotatebox{90}{\parbox[c]{1cm}{\centering Phys. \\ SC TPM}} &
    \begin{tabular}[c]{@{}c@{}}Availability\end{tabular} &
    \begin{tabular}[c]{@{}c@{}}Dedic.\\ HW\end{tabular} &
    \begin{tabular}[c]{@{}c@{}}Implementation\end{tabular} &
    \begin{tabular}[c]{@{}c@{}}Setup\end{tabular} &
    \begin{tabular}[c]{@{}c@{}}Attest.\end{tabular} &
    \begin{tabular}[c]{@{}c@{}}Comp.\end{tabular} \\
    \noalign{\hrule height 1pt} 
    ZK Proofs & \green{Null} & \green{P} & \green{P} & \green{P} & \green{P} & \green{P} & \green{SW} & \green{SW} & \green{Open Source} & \red{-- -- -- --} & \red{-- -- -- --} & \red{-- -- -- --} \\
    \noalign{\hrule height 1pt} 
    Nitro Enclaves & \red{Large} & \red{V} & \red{V} & \red{V} & \red{V} & \red{V} & \orange{AWS} & \red{Yes} & \red{Closed Source} & \red{-- -- --} & \green{+ +} & \green{+ +} \\
    \hline
    Arm TrustZone & \green{Small} & \red{V} & \red{V} & \red{V} & \red{V} & \red{V} & \green{High} & \red{Yes} & \red{HW Closed Source} & \green{+} & \green{+} & \red{--} \\
    \hline
    Intel SGX V1 & \green{Small} & \red{V} & \green{P} & \red{V} & \red{V} & \red{V} & \red{Deprecated} & \red{Yes} & \red{HW Closed-Source} & \red{-- --} & \green{+} & \red{-- --} \\
    \hline
    Intel SGX V2 & \green{Small} & \red{V} & \green{P} & \red{V} & \red{V} & \red{V} & \orange{Cloud} & \red{Yes} & \red{HW Closed-Source} & \red{-- --} & \green{+} & \green{+} \\
    \noalign{\hrule height 1pt} 
    AMD SEV & \red{Large} & \red{V} & \green{P} & \red{V} & \red{V} & \red{V} & \orange{Cloud} & \red{Yes} & \red{HW Closed-Source} & \red{-- -- --} & \green{+} & \green{+ +} \\
    \hline 
    Intel TDX & \red{Large} & \red{V} & \green{P} & \red{V} & \red{V} & \red{V} & \red{Coming} & \red{Yes} & \red{HW Closed-Source} & \red{-- -- --} & \green{+} & \green{+ +} \\
    \hline 
    ARM CCA & \red{Large} & \red{V} & \green{P} & \red{V} & \red{V} & \red{V} & \red{Coming} & \red{Yes} & \red{HW Closed-Source} & \red{-- -- --} & \green{+} & \green{+ +} \\
    \noalign{\hrule height 1pt} 
    TrustVisor\cite{mccune2010trustvisor} & \green{Small} & \red{V} & \red{V} & \red{V} & \red{V} & \red{V} & \red{Deprecated} & \green{No} & \green{Open Source} & \red{--} & \green{+ +} & \green{+ +} \\
    \hline
    Flicker~\cite{mccune2008flicker} & \green{Tiny} & \green{P} & \green{P} & \green{P} & \red{V} & \red{V} & \red{Deprecated} & \green{No} & \green{Open Source} & \red{-- -- --} & \red{-- -- --} & \red{-- -- --} \\
    \noalign{\hrule height 1pt} 
    \textbf{Argos} & \green{Small} & \green{P} & \green{P} & \green{P} & \red{V} & \red{V} & \green{High} & \green{No} & \green{Open Source} & \green{+} & \red{--} & \green{+ +} \\
    \noalign{\hrule height 1pt}
    \end{tabular}
    \vspace{10pt}
     \caption{Comparison of TEE Platforms across Security, Usability and Performance. \\ SC: Side Channel, \green{P: Protected}, \red{V: Vulnerable}. \change{\green{+}'s and \red{-}'s represent relative (and subjective) measure of performance}.
     }
    \label{tab:tee_platforms}
    \end{minipage}
\end{table*}

\section{Related Work}
\label{sec:relatedwork}

\noindent\textbf{TPMs and Hypervisor-based TEEs.}
Trusted platform modules (TPMs) combined with dynamic root-of-trust technologies have long served as dedicated hardware to enforce platform integrity.
Flicker~\cite{mccune2008flicker} proposes to use these two technologies combined to isolate and attest small pieces of application logic, or PALs while reducing the software TCB to a handful of lines of code.
No keys ever leave the TPM in Flicker, which gives it the same resistance to microarchitectural side channels as Argos (even if not discussed in the paper at the time).
However, their complete reliance on hardware mechanisms creates significant overheads, i.e., 2-3 orders of magnitude.
Virtualization technologies (making their appearance in processors around 2008) aim at isolating different trusted domains in the form of virtual machines (VMs).
They create a new privileged execution mode dedicated to the hypervisor and provide hardware primitives such as extended (or nested) page tables to enforce memory isolation between different security domains.
These technologies identified the opportunity to virtualize the TPM~\cite{perez2006vtpm} and provide confidentiality and integrity for trusted domains with minimal performance overhead.
TrustVisor~\cite{mccune2010trustvisor} is the first work to introduce this hypervisor-based architecture, which was eventually adopted by cloud providers such as AWS for their Nitro Enclaves~\cite{AWSNitroEnclaves}.
Virtual TPMs are also the standard for trusted computing in modern clouds~\cite{AzureVirtualTPM, raj2016ftpm, AWSNitroTPM}.
As a result, all these architectures are insecure under our threat model as they expose sensitive key material to microarchitectural side channels.

\smallskip
\noindent\textbf{Hardware TEE Platforms.}
All commercial hardware TEE platforms are vulnerable to microarchitectural side channels.
This includes, but is not limited to, Intel SGX~\cite{costan2016intel}, Intel TDX~\cite{cheng2024intel}, AMD SEV-SNP~\cite{sev2020strengthening}, ARM TrustZone~\cite{pinto2019demystifying} and ARM CCA~\cite{li2022design}.
\change{Secrets and key material are sometimes stored on secure co-processors, such as Intel ME~\cite{IntelCSME}, AMD PSP~\cite{buhren2019insecure}, or the Apple secure enclave~\cite{appleplatformsecurity}.
However, to our knowledge, all existing designs expose some secret material on the CPU such as platform-level secrets for Intel SGX/TDX~\cite{costan2016intel}, vTPM key material for AMD SEV-SNP~\cite{antonino2023flexible} or application-level keys for Apple Private Cloud Compute~\cite{ApplePrivateCloudSecurityGuide}.
}
Beyond their vulnerability to side channels, commercial platforms also suffer from availability issues as they require dedicated hardware and are not available on consumer-grade processors.
For example, Intel announced that SGX would be deprecated on non-server-grade processors~\cite{IntelSGXDeprecated} and also deprecated its remote attestation service~\cite{IntelSGXEPIDDeprecated}.
A comprehensive comparison of existing platforms with Argos can be found in \Cref{tab:tee_platforms}.
Academic platforms~\cite{costan2016sanctum, ferraiuolo2017komodo, lee2020keystone,  bahmani2021cure, feng2021scalable, azab2011sice, brasser2019sanctuary, deng2014equalvisor, feng2021scalable, van2023cheri, weiser2019timber} are limited in terms of how much hardware customization they can perform to defend against microarchitectural side channels.
As a result, the vast majority simply consider side channels out of scope, with the exception of Sanctum~\cite{costan2016sanctum} and MI6~\cite{bourgeat2019mi6} which are built on top of an open-source RISCV processor.

\smallskip
\noindent\textbf{Combining Cryptography and TEEs.}
Previous work has identified that protecting remote attestation mechanisms alone against microarchitectural side channels was easier than enforcing confidentiality of entire enclave programs~\cite{tramer2017sealed}.
Some even suggest to combine trusted hardware with cryptography to build hybrid mechanisms, but all show important performance overheads, use (insecure) SGX and are vulnerable to microarchitectural side channels.
CrypTFlow~\cite{kumar2020cryptflow} looks at secure inference and proposes to build maliciously-secure MPC from semi-honest MPC schemes using SGX.
They show a 3$\times$ performance overhead over the semi-honest scheme.
Chex-Mix~\cite{natarajan2021chex} leverages SGX with FHE to build secure inference, but considers microarchitectural side channels out of scope and even relies on SGX confidentiality for the privacy of the model weights.
They show a 142\% performance overhead compared to semi-honest FHE and consider key recovery attacks out of scope.
Viand et. al.~\cite{viand2023verifiable} introduces many useful formalisms that inspired Argos.
Their paper primarily focuses on combining FHE with cryptographic proofs but they also
consider an FHE-in-TEE approach and design a special protocol to optimize its performance.
Nevertheless, they use SGX, do not adapt their formalism to the FHE-in-TEE approach, and their results show an 80$\times$ performance overhead when compared to Argos.
In contrast, Argos presents a new TEE design for malicious and verifiable FHE with a dedicated remote attestation scheme. Argos is secure \emph{by design} against microarchitectural side channels and shows minimal performance overhead compared to semi-honest baselines.

%% file: sections/conclusion.tex
\section{Conclusion}
We present Argos, the first \emph{integrity-only} enclave platform designed to build maliciously-secure verifiable FHE.
Argos is secure \emph{by design} against microarchitectural side channels and transient execution attacks and can be used to build fully malicious and authenticated PSI and PIR schemes. It requires no specialized hardware and is compatible with commodity processors from 2008 onward. Argos only incurs 3\% average performance overhead for FHE computation, less than 8\% performance overhead for complex protocols, and no communication overhead. By demonstrating how to combine cryptography with trusted hardware, Argos paves the way for widespread deployment of FHE-based protocols beyond the semi-honest setting, without the overhead of cryptographic proofs. 

%% file: sections/appendix.tex
\appendix

\section{Primer on Measured Boot}
\label{app:primertpm}
Measured boot is obtained using a trust chain that binds the measurement of the operating system or hypervisor to a component that is inherently trusted by the system (the root).
Starting from the root, every element of the chain measures (hashes) and signs the binary of the next element, ensuring its integrity.
This process is sometimes referred to as secure boot, as, historically, the chain of trust was composed of the entire boot chain.
The root of trust (RoT) is the base case of this recursive security argument, often a hardware mechanism (e.g. read-only memory) that enforces integrity for the first instructions executed at boot and which integrity is supported by a certificate signed by the manufacturer~\cite{ermolov2016safeguarding, zimmer2016establishing}.

When elements of the boot chain are measured (for instance, before any memory can be trusted), these measurements need to be stored in a secure location where they cannot be tampered with, for instance, by a compromised BIOS.
To provide that functionality, a small chip or coprocessor, called a trusted platform module (TPM) will be in charge of securely storing hash-based measurements in special registers (PCRs).
The interface exposed by the TPM is quite limited. For example, measurements can only be hash-extended (never reset).
TPMs also have other capabilities, like long-term secrets storage referred to as \emph{secret sealing} (e.g., a key for disk encryption~\cite{tan2020deep}) that can only be accessed if the PCRs are in a specific state, i.e., the correct system has been booted.
The TPM also has attestation capabilities, which will use a private key tied to the TPM to sign or \emph{attest} the values of the PCRs and the state of the system.
However, the TPM is merely a subordinate to the RoT and the software in the chain of trust, as it needs to be instructed and fed inputs to extend its internal measurements.

Any code measured through that process, up to the final domain, is considered part of the trusted code base (TCB).
Code in the TCB is critical for the security of the system (it enforces the integrity of the next step in the chain of trust), and as a result, code in the TCB is expected to be bug-free.
This assumption is practically impossible to enforce for a TCB that contains millions of lines of code (e.g., the Linux kernel with 28MLOC), hence the incentive to exclude as much code as possible from the TCB.
A dynamic root of trust (DRoT) makes it possible to remove elements of the boot chain from the chain of trust, hence reducing the TCB.
DRoT is implemented as a hardware extension (e.g., Intel TXT or AMD DRMT~\cite{futral2013intel, nie2007dynamic}) that can instruct and control the TPM independently of any firmware or software.
However, DRoT and TPM were originally conceived to attest to one trusted domain (the OS).
That means 1) there is no mechanism to isolate more than one domain at a time 2) TPM hardware resources (PCRs, key storage, bandwidth) are sized for one domain and extremely constrained.

\section{Primer on Microarchitectural and Physical Attacks}
\label{app:primerattacks}
\subsection{Side Channel Attacks}
In side-channel attacks, an attacker leverages some shared state (initially not meant for communication) to learn information about the execution of a victim program and extract victim secrets.
There exist many types of side channels, and they can first be categorized by the medium or shared state that is exploited by the attacker to extract information.
For example, classic \textit{timing} side channels exploit the fact that the program execution time might vary depending on the value of some secret inputs.
It was shown early on~\cite{almeida2016verifying} that cryptographic algorithms such as RSA using \textit{non-constant-time} implementation (such as square and multiply) were vulnerable to these types of side channels.
Beyond simple timing, other types of side channel are
\begin{itemize}
    \item \textit{microarchitectural} side channels, which exploit shared \textit{microarchitectural} state. These include memory caches \cite{gotzfried2017cache, brasser2017software}, translation look-aside buffers (TLBs) \cite{gras2018translation}, branch predictors \cite{evtyushkin2018branchscope, aciiccmez2006predicting}, DRAM controllers \cite{wang2014timing}, or any other shared microarchitecture.
    \item \textit{physical} side channels, which exploit \textit{physical} mediums such as power consumption~\cite{wang2022hertzbleed, Lipp2021Platypus}, electromagnetic emissions~\cite{agrawal2003side}, or acoustic noise~\cite{backes2010acoustic}.
\end{itemize}
Finally, another way to classify \textit{physical} side channels is (1) if the attack can be mounted in \textit{software} or (2) if the attacker requires \textit{physical} access to the machine running the victim program.
Argos defend against all types of side channels targeting the CPU and is vulnerable to \textit{physical} side-channels targeting the TPM.
However, to our knowledge, no such attack has been published~(see Section \ref{sec:discussion}).

\subsection{Cache Side Channel Attacks}
Cache side channels are an example of \textit{microarchitectural} side channels. Here, the shared channel is shared caches. The caches are between the CPU and main memory (see fig. \ref{fig:computing_stack}) and ensure that the performance penalty of accessing a location in memory (100 cycles) can be amortized if the same memory value has been recently accessed.
We usually talk about a cache hierarchy, as a CPU usually contains a core-private fast (1 cycle) (but small) ``L1'' cache which is connected to other larger (but slower, 10 - 100 cycles) caches (``L2'' or ``L3'') closer to memory.
In principle, only the last level cache (LLC) is shared among cores, but this is machine-dependent.
That means that caches can be shared between a victim and an attacker program or \textit{temporarily} (between context switches on the same core~\cite{yarom2014flush+}) or \textit{spatially} (across cores if the cache is shared~\cite{liu2015last}).
To exploit caches as side channels, the attacker usually exploits the timing difference of loading a cache and non-cached value from memory.
We will describe (a simplified version) or Prime\&Probe~\cite{osvik2006cache}, a powerful cache side channel attacks:
\begin{enumerate}
    \item \textbf{PRIME:} The attacker will access a large array to fill up the cache with its own data.
    \item \textbf{WAIT:} The attacker will wait for the victim to run and access the shared cache.
    \item \textbf{PROBE:} The attacker will access the large array again, but time each memory access.
\end{enumerate}
If a memory access takes longer, that means that the corresponding value has been evicted from the cache by the victim when running, and the attacker can infer information regarding the \textit{addresses} accessed by the victim in memory.
In fact, to remove a piece of data from a cache entry, the new memory value must have the same \textit{set index}, hence sharing some common address bits with the original entry.
These attacks are efficient, but require the victim program to access memory locations whose addresses depend on a secret.
This is exactly why constant-time programming for cryptographic algorithms does not perform any memory accesses on secret-dependent addresses.

\subsection{Spectre Attacks}
Miroarchitectural side channel attacks are limited by the presence of gadgets in the victim code that can be exploited (e.g., secret dependent memory accesses).
These gadgets might exist in the code but may never be executed under ``normal'' sequential program execution.
Unfortunately, attacks from the Spectre family~\cite{kocher2018spectre, koruyeh2018spectre, chen2019sgxpectre} have demonstrated that such gadgets could be accessed \textit{speculatively} and weaponized into \textit{universal read gadgets}.

Speculative execution (sometimes called transient execution) is due to a family of speculative mechanisms for performance optimizations present on modern processors.
These include the pattern history table, the branch target buffer, and the return stack buffer, which all have their corresponding Spectre variants.
They make it possible to \textit{speculatively} execute code or load data before knowing if it corresponds to the correct control / data flow for the program.
If the microarchitecture guesses right, a lot of precious CPU cycles have been saved.
If the microarchitecture guesses wrong, some performance penalty needs to be paid to rollback the \textit{architectural} state (but not the \textit{microarchitectural} state) before correct execution can restart.
These optimizations are in large part responsible for the great performance of modern CPUs.
Disabling them altogether would have disastrous performance effects (200-300\% slowdown~\cite{bourgeat2019mi6}).
However, they make reasoning about the security of programs especially tricky as there is no clear \textit{speculative} execution flow for a given program.
In addition, when a speculative execution path is deemed wrong, the different pieces of microarchitectural states that were modified (e.g., cache) are usually not rolled back.
This has significant security implications, as this potentially secret-dependent state can now be observed by an attacker using adequate microarchitectural side channels.
Argos defend against these attacks by eliminating all microarchitectural side channels.

\subsection{Spectre V1}
As explained above, a large number of Spectre variants can be generated by combining a mechanism that triggers speculation with microarchitectural side channels.
We will now describe the original Spectre attack (V1~\cite{kocher2018spectre}) that exploits the branch predictor (pattern history table) and a cache side channel.
The branch predictor is one of the most common microarchitectural mechanisms for speculation.
When the CPU encounters a branch (i.e., \texttt{if}, \texttt{while}, or \texttt{for} statement), and before the result of the branch condition is known, the branch predictor guesses the direction of the branch (taken vs. non-taken) and speculatively executes the corresponding execution path.
A branch predictor is \textit{trained} using the history of correct directions for the branch to make its guesses more efficient.
That means that for a simple branch predictor, if a branch has been taken a dozen times in the same execution context, it will be predicted as taken in the next run.

Let us now describe the classic Spectre V1 attack.
Let us consider the following pseudo-code located in the victim program:
\begin{lstlisting}[language=C]
if(i > 0 && i < 10) {
    s = array1[i];
    b = array2[s];
}
\end{lstlisting}
Here, we assume that the attacker can call this code snippet by interacting through the victim program API.
Note that double-memory access is a perfect \textit{transmitter} for a cache side channel, as it accesses a memory value \texttt{s} and then leaks its content by accessing an address that depends on it (\texttt{array2[s]}).
However, thanks to the bound check on index \texttt{i}, the attacker should not be able to access arbitrary data location in victim memory and load it into \texttt{s}.
An attacker can exploit the branch predictor by first calling the code snippet repeatedly with inbound indices, training it to predict the branch on line 1 as non-taken. Subsequently calling the snippet with an out-of-bounds index causes the branch predictor to \textit{incorrectly} speculate, executing both the arbitrary memory access on line 2 and the cache side channel transmission on line 3. This creates a \textit{universal read gadget}, which enables arbitrary memory access within the victim's address space.

\subsection{Cold Boot}
Cold boot is a physical attack that requires physical access to the target machine running the victim program.
DRAM is volatile, but data values may persist for a few seconds / minutes even after a power switch-off.
If an attacker acts swiftly, it has the opportunity to \textit{cold boot} the machine with an attacker-controlled operating system (on a USB disk, for instance) and dump all the memory.
Argos defends against these attacks by never storing sensitive data in main memory.

\change{
\subsection{Fault Injection Attacks}
Fault injections are a different family of attacks than side channels, as they do not directly aim at extracting victim secrets, but rather at \textit{modifying} the victim control flow or its state by introducing faults into the system. 
Typically, such attacks require physical access to the system. 
Voltage glitching attacks are an example of such attacks that consist of lowering the power supply of a system in order to introduce glitches or faults into a system.
When properly synchronized and timed, changes in the power supply can affect the propagation of electric currents in the processor and affect the control or data flow of a program.
For instance, instructions can be skipped and the result of instructions can be affected, resulting in faulty outputs.
Other classic attacks include ``voltage/clock glitching, electromagnetic pulses, and laser-based attacks''~\cite{breier2022practical}.
Argos does not protect against such attacks.
However, in some cases, they can be mitigated in software by adding extra checks on the control flow.

\subsection{Rowhammer}
Rowhammer~\cite{zhang2024sok} is a \textit{software-mounted} fault injection attack that aims to modify the contents of the victim's memory. 
It exploits undesirable side effects in modern DRAM due to electromagnetic interference between closely located memory cells.
When repetitively activating a memory row (hammering the row), adjacent rows might leak charge from their cells due to the interference and some of their bit values might flip, even if never directly accessed by the program.
However, these attacks require the attacker to be able to access a memory row adjacent to the victim.
Moreover, such attacks can be mitigated by Argos by ensuring that different security domains use distinct memory partitions~\cite{Siloz}.
}

\section{Formal Definitions}
\noindent We use the formal definitions from Viand et. al.~\cite{viand2023verifiable}.

\subsection{FHE Properties}
\label{app:fheprop}

\noindent\textbf{Correctness}
A scheme is correct if any honest computation will decrypt to the expected result. More formally, a scheme is correct if for all functions $f$, and for all $x$ in the domain of $f$:
$$
\Pr\left[\Epsi.\text{Dec}_{\mathsf{sk}}(c_y) = f(x) \;\middle|\;
\begin{array}{l}
    (\mathsf{pk}, \mathsf{sk}) \leftarrow \Epsi.\text{Gen}(1^\lambda) \\
    c_x \leftarrow \Epsi.\text{Enc}_{\mathsf{key}}(x) \\
    c_y \leftarrow \Epsi.\text{Eval}_{\mathsf{key}}(c_x, f)
\end{array}
\right] = 1.
$$
\\

\noindent\textbf{Security}
We extend the definition of IND-CPA to our notion.
Note that we do not require an evaluation oracle since $\mathsf{Eval}_{\mathsf{pk}}$ 
is public and can be executed by the adversary directly. 
Formally, a scheme is secure if for any PPT adversary $\mathcal{A}$ 
and any function $f$ the advantage
$\mathsf{Adv}^{\mathsf{IND}\text{-}\mathsf{CPA}}[\mathcal{A}](\lambda) = 
2\left|\Pr[b = \hat{b}] - \frac{1}{2}\right|$ 
of the attacker in the following game is 
negligible in the security parameter $\lambda$:
$$
\begin{array}{l}
    \underline{\text{IND-CPA \textit{for} vFHE}} \\
    (\mathsf{pk}, \mathsf{sk}) \leftarrow \Epsi.\mathsf{Gen}(1^\lambda) \\
    (\mathbf{m}_0, \mathbf{m}_1) \leftarrow \mathcal{A}^{\mathcal{O}_{\Epsi.\mathsf{Enc}}}(1^\lambda, f, \mathsf{pk}) \\
    (c^*, \pi^*) \leftarrow \Epsi.\mathsf{Enc}_{\mathsf{pk}}(\mathbf{m}_b) \\
    \hat{b} \leftarrow \mathcal{A}^{\mathcal{O}_{\Epsi.\mathsf{Enc}}}(c^*).
\end{array}
$$

\noindent\textbf{Extension for $\epsilon$-approximate FHE schemes}
Approximate FHE schemes can be captured by changing the correctness property to the following:
A scheme is correct if for all functions
$f$, and for all $x$ in the domain of $f$:
$$
\Pr\left[\left|\left|\text{Dec}_{\text{sk}}(c_y) - f(x)\right|\right| \leq \varepsilon \;\middle|\;
\begin{array}{l}
    (\mathsf{pk}, \mathsf{sk}) \leftarrow \Epsi.\text{Gen}(1^\lambda) \\
    c_x \leftarrow \Epsi.\text{Enc}_{\mathsf{key}}(x) \\
    c_y \leftarrow \Epsi.\text{Eval}_{\mathsf{key}}(c_x, f)
\end{array}
\right] = 1
$$

\noindent where $\left|\left|\cdot\right|\right|$ is a scheme-specific norm and $\epsilon$ is a scheme-specific upper bound on the decoding error (which may
depend on $f$, pk, or other quantities of the scheme).
We leave out details on how to adapt our other definitions and proofs to this setup.

\subsection{Remote Attestation Properties}
\label{app:raprop}

\noindent\textbf{Completeness}
A scheme is complete if Verify will always accept an honestly computed result. More formally, a scheme is complete
if for all functions $g$, and for all $x$ in the domain of $g$:
$$
\Pr\left[\Pi.\text{Verify}_{\mathsf{sk}}(y, x, \pi_{y}) = \texttt{true} \;\middle|\;
\begin{array}{l}
    (\mathsf{pk}, \mathsf{sk}) \leftarrow \Pi.\text{Gen}(1^\lambda) \\
    (y, \pi_{y}) \leftarrow \Pi.\text{Prove}_{sk}(x, g)
\end{array}
\right] = 1.
$$
\\

\noindent\textbf{Soundness}
A scheme is sound if the adversary cannot make Verify
accept an incorrect answer. Formally, a scheme is sound if
for any PPT adversary and any function $g$ the following
probability is negligible in the security parameter $\lambda$:
$$
\Pr\left[
\begin{array}{c}
    \Pi.\text{Verify}_{\mathsf{sk}}(y, x, \pi_{y}) = 1 \\
    \land \\
    y \neq g(x)
\end{array}
\;\middle|\;
\begin{array}{l}
    (\mathsf{pk}, \mathsf{sk}) \leftarrow \Pi.\text{Gen}(1^\lambda) \\
    x \leftarrow \mathcal{A}(\mathsf{pk}) \\
    \text{x is in the domain of g}\\
    (y, \pi_{y}) \leftarrow \mathcal{A}(\mathsf{pk}, x)
\end{array}
\right].
$$

\subsection{vFHE Properties}
\label{app:vfheprop}

\noindent\textbf{Correctness}
A scheme is correct if any honest computation will decrypt to the expected result. More formally, a scheme is correct if for all functions $f$, and for all $x$ in the domain of $f$:
$$
\Pr\left[\text{Dec}_{\mathsf{sk}}(c_y) = f(x) \;\middle|\;
\begin{array}{l}
    (\mathsf{pk}, \mathsf{sk}) \leftarrow \mathsf{Gen}(1^\lambda) \\
    c_x \leftarrow \mathsf{Enc}_{\mathsf{key}}(x) \\
    (c_y, \pi_y) \leftarrow \mathsf{Eval}_{\mathsf{key}}(c_x, f)
\end{array}
\right] = 1.
$$
\\

\noindent\textbf{Completeness}
A scheme is complete if Verify will always accept an honestly computed result. More formally, a scheme is complete
if for all functions $f$, and for all $x$ in the domain of $f$:
$$
\Pr\left[\mathsf{Verify}_{\mathsf{sk}}(c_y, c_x, \pi_y) = 1 \;\middle|\;
\begin{array}{l}
    (\mathsf{pk}, \mathsf{sk}) \leftarrow \mathsf{Gen}(1^\lambda) \\
    c_x \leftarrow \mathsf{Enc}_{\mathsf{key}}(x) \\
    (c_y, \pi_y) \leftarrow \mathsf{Eval}_{\mathsf{key}}(c_x, f)
\end{array}
\right] = 1.
$$
\\

\noindent\textbf{Soundness}
A scheme is sound if the adversary cannot make Verify
accept an incorrect answer. Formally, a scheme is sound if
for any PPT adversary $\mathcal{A}$ and any function $g$ the following
probability is negligible in the security parameter $\lambda$:
$$
\Pr\left[
\begin{array}{c}
    \mathsf{Verify}_{\mathsf{sk}}(c_y, c_x, \pi_{y}) = 1 \\
    \land \\
    c_y \neq \Epsi.\text{Eval}(c_x, f)
\end{array}
\;\middle|\;
\begin{array}{l}
    (\mathsf{pk}, \mathsf{sk}) \leftarrow \mathsf{Gen}(1^\lambda) \\
    x \leftarrow \mathcal{A}^{\mathcal{O}_{\mathsf{Enc}}, \mathcal{O}_{\mathsf{Dec}}}(\mathsf{pk}) \\
    c_x \leftarrow \mathsf{Enc}_{\mathsf{key}}(x) \\
    (c_y, \pi_y) \leftarrow \mathcal{A}^{\mathcal{O}_{\mathsf{Enc}}, \mathcal{O}_{\mathsf{Dec}}}(c_x)
\end{array}
\right].
$$
Note that in the literature~\cite{viand2023verifiable, atapoor2024verifiable}, soundness is sometimes defined using the conditional probability of
$$\mathsf{Verify}_{\mathsf{sk}}(c_y, c_x, \pi_{y}) = 1 \quad \land \quad \Epsi.\text{Dec}(c_y) \neq f(x).$$
This definition is impractical as it allows an attacker to compute a different circuit that represents the same function.
We use a more restrictive definition instead, even if it requires the $\Epsi.\text{Eval}$ algorithm to be deterministic (which it is in our case).
\\

\noindent\textbf{Security}
We extend the definition of IND-CCA1 to our notion. Note 
that we do not require an evaluation oracle since $\mathsf{Eval}_{\mathsf{pk}}$ 
is public and can be executed by the adversary directly. 
Formally, a scheme is secure if for any PPT adversary $\mathcal{A}$ 
and any function $f$ the advantage
$\mathsf{Adv}^{\mathsf{IND}\text{-}\mathsf{CCA1}}[\mathcal{A}](\lambda) = 
2\left|\Pr[b = \hat{b}] - \frac{1}{2}\right|$ 
of the attacker in the following game is 
negligible in the security parameter $\lambda$:
$$
\begin{array}{l}
    \underline{\text{IND-CCA1 \textit{for} vFHE}} \\
    (\mathsf{pk}, \mathsf{sk}) \leftarrow \mathsf{Gen}(1^\lambda) \\
    (\mathbf{m}_0, \mathbf{m}_1) \leftarrow \mathcal{A}^{\mathcal{O}_{\mathsf{Enc}},\mathcal{O}_{\mathsf{Dec}}}(1^\lambda, f, \mathsf{pk}) \\
    (c^*, \pi^*) \leftarrow \mathsf{Enc}_{\mathsf{pk}}(\mathbf{m}_b) \\
    \hat{b} \leftarrow \mathcal{A}^{\mathcal{O}_{\mathsf{Enc}}}(c^*)
\end{array}
$$

\section{Extended Proof: Circuit-Level vFHE}
\label{app:securityfullcirtuitlevelfhe}

\noindent\textbf{Construction}
Let us consider the vFHE scheme (Gen, Enc, Eval, Verify, Dec) described in Section \ref{sec:defvfhe}.
The construction is as follows: we take an FHE scheme ($\Epsi.\text{Gen}$, $\Epsi.\text{Enc}$, $\Epsi.\text{Eval}$, $\Epsi.\text{Dec}$) and a remote attestation scheme ($\Pi.\text{Gen}$, $\Pi.\text{Prove}$, $\Pi.\text{Verify}$).

\noindent We then have
\begin{itemize}
    \item $\text{Gen}(1^\lambda) \rightarrow ((\text{pk}_{\Epsi}, \text{sk}_{\Epsi}), (\text{pk}_{\Pi}, \text{sk}_{\Pi}))$ \\ with $(\text{pk}_{\Epsi}, \text{sk}_{\Epsi}) = \Epsi.\text{Gen}(1^\lambda)$ and $(\text{pk}_{\Pi}, \text{sk}_{\Pi}) = \Pi.\text{Gen}(1^\lambda)$
    \item $\text{Enc}(x) \rightarrow c_x$ with $c_x = \Epsi.\text{Enc}(x)$
    \item $\text{Eval}(c_x, f) \rightarrow (c_y, \pi_{y})$ with ($c_y, \pi_{y}) = \Pi.\text{Prove}(c_x, \Epsi.\text{Eval}(~.~, f))$
    \item $\text{Verify}(c_y, c_x, \pi_{y}) \rightarrow b$ with $b = \Pi.\text{Verify}(c_y, c_x, \pi_{y})$.
    \item $\text{Dec}(c_y) \rightarrow y$ with $y = \Epsi.\text{Dec}(c_y)$
\end{itemize}

\bigskip

\noindent\textbf{Correctness}
Correctness of $\text{Dec}_{sk}$ follows immediately from the correctness of the underlying FHE scheme. 
Specifically, we have that
$$
\Pr\left[\Epsi.\text{Dec}_{\mathsf{sk}}(c_y) = f(x) \;\middle|\;
\begin{array}{l}
    (\mathsf{pk}, \mathsf{sk}) \leftarrow \Epsi.\text{Gen}(1^\lambda) \\
    c_x \leftarrow \Epsi.\text{Enc}_{\mathsf{key}}(x) \\
    c_y \leftarrow \Epsi.\text{Eval}_{\mathsf{key}}(c_x, f)
\end{array}
\right] = 1.
$$
and so, rewriting the terms, it follows that 
$$\text{Dec}_{sk}(c_y) = \Epsi.\text{Dec}_{sk}(c_y) = f(x)$$
with probability $1$.
\\

\noindent\textbf{Completeness}
Completeness of $\text{Verify}_{sk}$ follows immediately from the completeness of the underlying remote attestation scheme. 
Specifically, we have that
$$
\Pr\left[\Pi.\text{Verify}_{\mathsf{sk}}(y, x, \pi_{y}) = \texttt{true} \;\middle|\;
\begin{array}{l}
    (\mathsf{pk}, \mathsf{sk}) \leftarrow \Pi.\text{Gen}(1^\lambda) \\
    (y, \pi_{y}) \leftarrow \Pi.\text{Prove}_{sk}(x, g)
\end{array}
\right] = 1
$$
and so, rewriting the terms, it follows that 
$\mathsf{Verify}_{\mathsf{sk}}(c_y, c_x, \pi_y) = \Pi.\text{Verify}_{\mathsf{sk}}(c_y, c_x, \pi_{y}) = \texttt{true}$ with probability $1$.
\\

\noindent\textbf{Security}
We prove the security of the scheme via a reduction to the security of the underlying FHE scheme.
Consider a probabilistic polynomial time (PPT) attacker \Att{} with oracle access to Enc and Dec that breaks the semantic security of our vFHE scheme.
Formally, this means that \Att{} can produce two messages $m_0$ and $m_1$ such that it can distinguish with non-negligible advantage between $\text{Enc}(m_0)$ and $\text{Enc}(m_1)$.
We show that we can build a PPT attacker $\mathcal{B}$ with only oracle access to $\Epsi.\text{Enc}$ that runs $\mathcal{A}$ to contradict the semantic security of the FHE scheme with the same advantage. 

It is sufficient to show that $\mathcal{B}$ can emulate a Dec oracle for the vFHE scheme without having access to a $\Epsi.\text{Dec}$ oracle for the underlying FHE scheme.
$\mathcal{B}$ emulates the Dec oracle by executing the following sequence of steps:
\begin{enumerate}
    \item
        For each query to Enc issued by $\mathcal{A}$, $\mathcal{B}$ makes an oracle request to its $\Epsi.\text{Enc}$ oracle and stores the latest $x$ used and the $c_x$ obtained.
        It response to $\mathcal{A}$ with $c_x$.
    \item 
        Then, when $\mathcal{A}$ sends an oracle request to $\mathcal{O}_{\text{Dec}}$ with a ciphertext $c_y$ (along with $\pi_{y}$), $\mathcal{B}$ runs $\Pi.\text{Verify}(c_y, c_x, \pi_{y})$ (which is a polynomial algorithm) by using the latest $c_x$ it stored in the previous step.
    \begin{itemize}
        \item 
            If $\Pi.\text{Verify}(c_y, c_x, \pi_{y}) = \texttt{true}$, then $\mathcal{B}$ uses the stored value of $x$ to compute $f(x)$ ($f$ is a finite circuit) and sends it to $\mathcal{A}$.

        \item 
            Otherwise, $\mathcal{B}$ responds with $\bot$.
    \end{itemize}

    \item 
        When $\mathcal{A}$ sends the challenge messages $(m_0, m_1)$, $\mathcal{B}$ forwards them to its own challenger as the challenge tuple. 
    \item 
        After the challenge, $\mathcal{B}$ continues to respond to the Enc oracle as before. 

    \item 
        Eventually, $\mathcal{B}$ outputs as $\mathcal{A}$ does. 
\end{enumerate}

Because the underlying remote attestation scheme is sound, $\Pi.$Verify will only return \texttt{true} (with overwhelming probability) if $c_y = \Epsi.\text{Eval}(c_x, f)$, and because the underlying FHE scheme is correct, $\text{Dec}({c_y}) = f(x)$ with high probability. 
As a result, $\mathcal{O}_{\text{Dec}}$ is exactly a Dec oracle for the vFHE scheme.
That means $\mathcal{B}$ can successfully emulate the required oracles and use $\mathcal{A}$ to distinguish with non-negligible advantage between $\text{Enc}(m_0)$ and $\text{Enc}(m_1)$.
\\

\noindent This contradicts the security assumption of the underlying FHE scheme, concluding the proof. 
\\

\noindent\textbf{Soundness}
We now turn to analyzing the soundness property.
We prove the soundness of the scheme via a reduction to the soundness of the underlying remote attestation scheme.
We make use of a hybrid argument (changes are \textcolor{blue}{highlighted}). 
\begin{itemize}
    \item \textbf{H0:}
        For any PPT adversary $\mathcal{A}$ and any function $g$ the following
    probability is negligible in the security parameter $\lambda$:
    $$
\Pr\left[
\begin{array}{c}
    \mathsf{Verify}_{\mathsf{sk}}(c_y, c_x, \pi_{y}) = 1 \\
    \land \\
    c_y \neq \Epsi.\text{Eval}(c_x, f)
\end{array}
\;\middle|\;
\begin{array}{l}
    (\mathsf{pk}, \mathsf{sk}) \leftarrow \mathsf{Gen}(1^\lambda) \\
    x \leftarrow \mathcal{A}^{\mathcal{O}_{\mathsf{Enc}, \mathcal{O}_{\mathsf{Dec}}}(\mathsf{pk})} \\
    c_x \leftarrow \mathsf{Enc}_{\mathsf{key}}(x) \\
    (c_y, \pi_y) \leftarrow \mathcal{A}^{\mathcal{O}_{\mathsf{Enc}}, \mathcal{O}_{\mathsf{Dec}}}(c_x)
\end{array}
\right].
$$
    \item \textbf{H1:}
        For any PPT adversary $\mathcal{A}$ and any function $g$ the following
    probability is negligible in the security parameter $\lambda$:
    $$
\Pr\left[
\begin{array}{c}
    \mathsf{Verify}_{\mathsf{sk}}(c_y, c_x, \pi_{y}) = 1 \\
    \land \\
    c_y \neq \Epsi.Eval(c_x, f)
\end{array}
\;\middle|\;
\begin{array}{l}
    (\mathsf{pk}, \mathsf{sk}) \leftarrow \mathsf{Gen}(1^\lambda) \\
    x \leftarrow \textcolor{blue}{\mathcal{A}^{\mathcal{O}_{\mathsf{Dec}}}(\mathsf{pk})} \\
    c_x \leftarrow \mathsf{Enc}_{\mathsf{key}}(x) \\
    (c_y, \pi_y) \leftarrow \textcolor{blue}{\mathcal{A}^{\mathcal{O}_{\mathsf{Dec}}}(c_x)}
\end{array}
\right].
$$
    \item \textbf{H2:}
        For any PPT adversary $\mathcal{A}$ and any function $g$ the following
    probability is negligible in the security parameter $\lambda$:
    $$
\Pr\left[
\begin{array}{c}
    \mathsf{Verify}_{\mathsf{sk}}(c_y, c_x, \pi_{y}) = 1 \\
    \land \\
    c_y \neq \Epsi.Eval(c_x, f)
\end{array}
\;\middle|\;
\begin{array}{l}
    (\mathsf{pk}, \mathsf{sk}) \leftarrow \mathsf{Gen}(1^\lambda) \\
    x \leftarrow \textcolor{blue}{\mathcal{A}(\mathsf{pk})} \\
    c_x \leftarrow \mathsf{Enc}_{\mathsf{key}}(x) \\
    (c_y, \pi_{y}) \leftarrow \textcolor{blue}{\mathcal{A}(c_x)}
\end{array}
\right].
$$
    \item \textbf{H3:}
        For any PPT adversary $\mathcal{A}$ and any function $g$ the following
    probability is negligible in the security parameter $\lambda$:
$$
\Pr\left[
\begin{array}{c}
    \textcolor{blue}{\Pi.\text{Verify}_{\mathsf{sk}}(y, x, \pi_{y})} = 1 \\
    \land \\
    \textcolor{blue}{y \neq g(x)}
\end{array}
\;\middle|\;
\begin{array}{l}
    (\mathsf{pk}, \mathsf{sk}) \leftarrow \textcolor{blue}{\Pi.\text{Gen}(1^\lambda)} \\
    x \leftarrow \mathcal{A}(\mathsf{pk}) \\
    \textcolor{blue}{\text{x is in the domain of g}}\\
    (y, \pi_{y}) \leftarrow \textcolor{blue}{\mathcal{A}(\mathsf{pk}, x)}
\end{array}
\right].
$$
\end{itemize}

\noindent\textbf{H0 $\rightarrow$ H1:}
Let us assume there exists an attacker \Att{} with oracle access to, Enc, and Dec that can win the challenge described in H1 with non-negligible probability.
Because the attacker in H1 has access to the FHE scheme public key, it can emulate $\mathcal{O}_{\mathsf{Enc}}$ and leverage \Att{} to win H2 with non-negligible probability.
\\

\noindent\textbf{H1 $\rightarrow$ H2:}
Let us assume there exists an attacker \Att{} with oracle access to Enc and Dec that can win the challenge described in H1 with non-negligible probability.
Similarly to how we proved the security for our vFHE scheme, the attacker in H2 can emulate $\mathcal{O}_{\mathsf{Dec}}$ and leverage \Att{} to win H2 with non-negligible probability.
\\

\noindent\textbf{H2 $\rightarrow$ H3:}
H3 is exactly H2 after rewriting the terms using the notation from the remote attestation scheme, and choosing a specific family of functions for $g$ (functions in the form $\Epsi.Eval(., f)$ for any functions $f$). Because H3 should hold for any function $g$, a PPT adversary that can win H2 with non-negligible probability can also win H3 with non-negligible probability.
\\

Because H3 is exactly the soundness property of our underlying remote attestation scheme, this concludes our proof by reduction. 

\section{Security Argument: Auth. PIR}
\label{app:secargpir}
In this section, we sketch how Argos can be used to instantiate authenticated PIR. 
The first requirement towards building authenticated PIR is extending Argos to support public inputs. 

\subsection{Extending vFHE for Public Server Inputs}
\label{app:vfheextentionpub}
\noindent\textbf{Accepting public inputs.}
We extend our vFHE scheme to allow public server input $w$ for $f$ as follows. 
We define a function $h(\cdot)$ that takes $w$ as input and outputs $w$ if it is ``well formed'' or halts otherwise. In practice, this is equivalent to type-checking $w$.
We now define $\Epsi.\text{Eval}'(c_x, w, h, f)$ that evaluates $\Epsi.\text{Eval}(c_x, f(\cdot, h(w)))$ and return the output.

\bigskip

\noindent\textbf{Authenticating server inputs.}
To authenticate server input, we assume that the server commits to its public database ahead of time.
As in prior work on authenticated PIR, we assume that the client obtains this commitment \textit{out-of-band}.
We now define a new family of vFHE schemes defined as (Commit, Gen, Enc, Eval, Verify, Dec), which is augmented with the new Commit algorithm.
The server uses Commit to commit to its public input $w$ by computing $Commit(w) \leftarrow C(w)$.
Here, $C$ is a commitment scheme (e.g., a Merkle tree) that is \textit{computationally binding} (informally, the probability to find $x \neq x'$ so that $C(x) = C(x')$ is negligible) and because the input is public, we do not require the commitment scheme to be hiding.
The server then transmits this commitment to the client (e.g., by posting it to a public bulletin board).
\\

\noindent\textbf{Putting everything together.}
We define $\Pi.\text{Prove}'$ which emulates $\Pi.\text{Prove}$, but ensures that on top of a commitment to $\Epsi.\text{Eval}(c_x, f(\cdot, \cdot))$, the transcript $\pi_y$ also contains a commitment to $h(x)$ and $C(w)$. 
We use $\Pi.\text{Prove}'$ to define our new Eval' function.
\begin{itemize}
    \item $\text{Eval}'(c_x, f) \rightarrow \Pi.\text{Prove}'(c_x, \Epsi.\text{Eval}'(~.~, w, h, f))$.
\end{itemize}

\noindent We also define $\text{Verify}'$ which 
\begin{enumerate}
    \item takes $C(w)$ as input;
    \item checks that $C(w)$ matches the commitment included in $\pi_y$;
    \item \begin{itemize}
        \item \textit{if} commitments match, evaluate the original $\text{Verify}$ and return the output.
        \item \textit{else} return \texttt{false}.
    \end{itemize} 
\end{enumerate}
In summary, our new vFHE scheme is defined by (Commit, Gen, Enc, Eval', Verify', Dec).
Given these changes, adapting the vFHE properties is straightforward.
Using the binding property of the Commit algorithm, all the security games can be extended to include the initial execution of Commit by the adversary.
We omit providing the formal properties and proof, since this extension is straightforward to derive and verify.

\subsection{Semi-Honest PIR from FHE}
We now consider a simple FHE-based PIR scheme.
We note that this approach can be extended to specific PIR schemes (e.g.,~\cite{angel2018pir}) in natural ways, but for simplicity we restrict ourselves to the FHE-based approach.
More concretely, consider an FHE scheme ($\Epsi.\text{Gen}$, $\Epsi.\text{Enc}$, $\Epsi.\text{Eval}$, $\Epsi.\text{Dec}$).
We construct our PIR scheme by evaluating (using FHE) the query function $f(\cdot, w)$ that takes an input index $x$ and returns $w[x]$, i.e., the $x$th value in the server database.
We now sketch the two required properties for PIR schemes: correctness and privacy.
These will help define authenticated PIR in the next section. 
\\

\noindent\textbf{Correctness.} Informally, a PIR scheme is \emph{correct} if, when an honest client interacts with honest servers, the client always retrieves $w[x]$ (i.e., the $x$th value in the database $w$).
\\

\noindent\textbf{Privacy.} A PIR scheme satisfies \emph{privacy} if an honest-but-curious server does not learn information regarding $x$.
\\

\noindent\textbf{Security Argument (folklore).} It is trivial to check that the correctness and security of the PIR scheme constructed by having the server evaluate the circuit that computes $w[x]$ given an encryption of $x$ under FHE, reduces to the correctness and security of the underlying FHE scheme.

\subsection{Authenticated PIR from vFHE}
We now turn to constructing authenticated PIR from verifiable FHE. 

\bigskip

\noindent\textbf{Construction.}
We build an authenticated PIR scheme by using the same construction of PIR described above, but swapping out the FHE scheme with our vFHE scheme that supports authenticated public server input (see \Cref{app:vfheextentionpub}).

\bigskip 

\noindent\textbf{Authenticated PIR properties}
For authenticated PIR, we sketch the properties from Colombo et. al.~\cite{colombo2023authenticated}.
In the case of authenticated PIR, the security property is extended to support a malicious server, and a new integrity property is added. Correctness is the same.
\\

\noindent\textbf{Integrity.} An authenticated-PIR scheme preserves \emph{integrity} if, when an honest client interacts with a malicious server, the client either: outputs $w[x]$ if the server followed the protocol or outputs the error symbol $\bot$ if the server deviated from the protocol.
\\

\noindent\textbf{Privacy (against malicious servers).} An authenticated PIR scheme satisfies \emph{privacy} if a malicious server does not learn information regarding $x$, even if the servers learn whether the client's output was the error symbol $\bot$ during reconstruction.

\bigskip

\noindent\textbf{Security Argument.} 
We now briefly argue why the scheme using vFHE with authenticated public server input satisfies these properties. 
Correctness follows from the correctness of the PIR scheme. Integrity reduces to the soundness of the underlying vFHE scheme, while security reduces to the underlying CCA1 security of the vFHE scheme. We leave out the detailed proofs.

\subsection{Extensions to More Complex PIR Schemes}
Most FHE-based PIR schemes are not implemented as a simple FHE circuit and utilize other techniques to optimize performance (e.g., cuckoo hashing). 
These schemes fit into our formalism by extending $h(x)$ to also contain any server input reprocessing. Because $h$ is also part of the attested transcript, the resulting PIR scheme run using Argos is verifiable (and authenticated).

\section{Security Argument: Authenticated PSI}
\label{app:secargpsi}
We now turn to our second application: authenticated PSI.
Similarly to FHE-based PIR schemes, FHE-based PSI schemes can also be formalized as the evaluation of a specific circuit under FHE, albeit with a more complex pre-processing of the server input (in particular to enforce \textit{server privacy}).
This fits our formalism by extending $h(x)$ to contain this extra pre-processing of server input.
Similarly to authenticated PIR, we could show that our vFHE scheme is sufficient to enforce analogous properties for authenticated PSI (correctness, integrity, client privacy, even if PIR security is often defined using an ideal functionality).
We omit these proofs in the interest of space.
Instead, we focus on the main difference with PIR, which is ``server privacy.'' 
\\

\noindent\textbf{Server Privacy.}
Informally, for a given $x$, for two server inputs $w$ and $w'$ that result in the same output $y$, there should be negligible advantage for a PPT adversary to distinguish between $(c_y,\pi_y)$ and $(c_y',\pi_y')$.
\\

Because most FHE schemes are not circuit-private, the output $c_y$ can leak some information about the server's input to the secret key holder.
To achieve server privacy with regard to $c_y$, FHE-based PSI schemes pre-process and mask server input using an oblivious pseudo-random function (OPRF)~\cite{chen2018labeled}.
As explained earlier, this pre-processing is captured by $h(\cdot)$.
This construction was shown by previous work to be \textit{server private} with regard to $c_y$.
Because our vFHE scheme does not modify the underlying FHE scheme or PSI scheme, we inherit this property and only need to ensure that $\pi_y$ does not breach the privacy of the server.
Given our construction, it is sufficient for us to choose a commitment scheme $C$ that is \textit{hiding}.
Informally, $\pi_y$ contains a commitment to $\Epsi.\text{Eval}(c_x, f(\cdot, \cdot))$ and $h(x)$, which are independent of $w$, and $C(w)$, which is hiding.
As a result, our construction is \textit{server private} with regard to $(c_y, \pi_y)$ and combining a PSI scheme with Argos successfully instantiates an authenticated PSI scheme.